\documentclass[acmsmall]{acmart}

\AtBeginDocument{%
  \providecommand\BibTeX{{%
    \normalfont B\kern-0.5em{\scshape i\kern-0.25em b}\kern-0.8em\TeX}}}

\setcopyright{acmcopyright}
\copyrightyear{2018}
\acmYear{2018}
\acmDOI{10.1145/1122445.1122456}

\acmJournal{TACO}
\acmVolume{37}
\acmNumber{4}
\acmArticle{111}
\acmMonth{8}

\newcommand{\arch}{PIMBALL}
\usepackage{graphicx}
\usepackage{wrapfig}
\usepackage{subcaption}
\usepackage{siunitx}
\graphicspath{ {Images/} }
\begin{document}

\title{PIMBALL: Binary Neural Networks in Spintronic Memory}

\author{Salonik Resch}
\email{resc0059@umn.edu}
\orcid{1234-5678-9012}
\affiliation{%
  \institution{University of Minnesota, Twin Cities}
  \streetaddress{200 Union St SE}
  \city{Minneapolis}
  \state{Minnesota}
  \postcode{55455}
}

\author{S. Karen Khatamifard}
\email{khatami@umn.edu}
\affiliation{%
  \institution{University of Minnesota, Twin Cities}
  \streetaddress{200 Union St SE}
  \city{Minneapolis}
  \state{Minnesota}
  \postcode{55455}
}

\author{Zamshed Iqbal Chowdhury}
\email{chowh005@umn.edu}
\affiliation{%
  \institution{University of Minnesota, Twin Cities}
  \streetaddress{200 Union St SE}
  \city{Minneapolis}
  \state{Minnesota}
  \postcode{55455}
}

\author{Masoud Zabihi}
\email{zabih003@umn.edu}
\affiliation{%
  \institution{University of Minnesota, Twin Cities}
  \streetaddress{200 Union St SE}
  \city{Minneapolis}
  \state{Minnesota}
  \postcode{55455}
}

\author{Zhengyang Zhao}
\email{zhaox526@umn.edu}
\affiliation{%
  \institution{University of Minnesota, Twin Cities}
  \streetaddress{200 Union St SE}
  \city{Minneapolis}
  \state{Minnesota}
  \postcode{55455}
}

\author{Jian-Ping Wang}
\email{jpwang@umn.edu}
\affiliation{%
  \institution{University of Minnesota, Twin Cities}
  \streetaddress{200 Union St SE}
  \city{Minneapolis}
  \state{Minnesota}
  \postcode{55455}
}

\author{Sachin S. Sapatnekar}
\email{sachin@umn.edu}
\affiliation{%
  \institution{University of Minnesota, Twin Cities}
  \streetaddress{200 Union St SE}
  \city{Minneapolis}
  \state{Minnesota}
  \postcode{55455}
}

\author{Ulya R. Karpuzcu}
\email{ukarpuzc@umn.edu}
\affiliation{%
  \institution{University of Minnesota, Twin Cities}
  \streetaddress{200 Union St SE}
  \city{Minneapolis}
  \state{Minnesota}
  \postcode{55455}
}

\renewcommand{\shortauthors}{Resch, et al.}

\begin{abstract}

Neural networks span a wide range of applications of industrial and commercial significance. Binary neural networks (BNN) are particularly effective in trading accuracy for performance, energy efficiency or hardware/software complexity. Here, we introduce a spintronic, re-configurable in-memory BNN
accelerator, \arch: \textbf{P}rocessing \textbf{I}n \textbf{M}emory \textbf{B}NN \textbf{A}cce\textbf{L(L)}erator, which 
%\arch\
allows for massively parallel and energy efficient computation.  \arch\ is capable of being used
as a standard spintronic memory (STT-MRAM) array and a computational substrate simultaneously.  We evaluate
\arch\ using multiple image classifiers and 
a genomics kernel.  Our
simulation results show that \arch\ is more energy efficient than alternative
CPU, GPU, and FPGA based implementations while delivering higher throughput.

\end{abstract}

\begin{CCSXML}
<ccs2012>
<concept>
<concept_id>10010520.10010521.10010542</concept_id>
<concept_desc>Computer systems organization~Other architectures</concept_desc>
<concept_significance>300</concept_significance>
</concept>
</ccs2012>
\end{CCSXML}

\ccsdesc[300]{Computer systems organization~Other architectures}

\keywords{Processing in Memory, non-volatile memory, binary neural networks, computational random access memory}

\maketitle

\section{Introduction}
\label{sec:introduction}
\noindent Neural networks (NN) have gained renowned attention in solving a diverse set of recognition, classification or optimization problems. Thus, there is great incentive to improve the performance and energy efficiency by hardware specialization. NN algorithms tend to be both compute and memory intensive due to large input data sets and the large number of operations required to process each input. Typical network sizes are too large to fit on chip, and thus require a supporting memory structure (typically DRAM) which has a high time and energy cost. This computational and storage demand often relegates neural networks to server/cloud based services, given that they are taxing on massive hardware resources and are typically impractical for mobile platforms at the edge.

Data representation (of network parameters such as weights) dictates computational and storage complexity.
Pure floating-point heavy designs capable of handling main neural network operations~\cite{dadiannao, zhang2015optimizing} usually impose 
the same data representation on the entire network, while advanced training techniques can reveal the minimum bit-width for different layers~\cite{gupta2015deep,han2015deep}.
Weight sharing, where a single value can represent multiple weights in the network, can reduce (on-chip) storage complexity significantly~\cite{chen2015compressing}.
Pruning can also help by making weight matrices sparse. Sparse matrix algebra decreases the number of operations required by construction, and thereby can improve performance \cite{han2016eie,han2015learning}. 

Binary Neural Networks (BNN) represent a recent surprising add-on to the design space, which trade accuracy for efficiency by allocating only a single bit for each weight: 0 representing the value -1 and 1 representing the value +1. This significantly reduces the storage complexity. In comparison, fixed point representation of weights is typically 8 to 32 bits. Their great energy efficiency also makes BNNs a leading candidate for mobile applications, where conserving battery life is paramount. Furthermore, on the extreme end, machine learning can also be of use on disposable electronics \cite{conti2018xnor,manic2016intelligent}. This domain is even more forgiving to accuracy loss in exchange of extreme energy efficiency and much lower hardware complexity. Thus, BNNs are very promising  to bring machine learning to the emerging (ultra) low power applications.

Effective methods for BNN training exist, to render near state-of-the-art accuracy in the MNIST
(hand-written digit recognition), CIFAR-10 (image classification), and SVHN (street view digit recognition) datasets~\cite{BINARIZED}. Other BNN proposals
such as XNOR-Net \cite{xnornet} and DoReFa-Net \cite{dorefanet}, are based on AlexNet, and use the much larger ImageNet dataset for classification. BNNs do tend to suffer from a significant loss in accuracy for this larger dataset. However, techniques exist to compensate for the binarization and regain accuracy by selectively (and sparingly) re-introducing  non-binary operations \cite{conti2018xnor}. Thus, BNNs are suitable for larger applications as well, as long as the accompanied accuracy loss is tolerable.

Perhaps the most significant advantage of binarization is simplification of the
underlying hardware. During the forward pass -- which is at the core of inference but is also required for training, most arithmetic operations 
reduce to bit-wise operations \cite{BINARIZED}. For example, a multiplication simply becomes an XNOR. This translates into significant gains in energy and performance, as the most common operation in (B)NN computation is multiply-and-accumulate (MAC).  As a result, recent FPGA-based BNN accelerators \cite{FPBNN,FINN,fu2018towards} deliver faster and more energy efficient execution than CPU- and GPU-based counterparts.

Many of the emerging Processing-In-Memory (PIM) solutions, that 
can perform bit-wise operations inside the memory array   {\cite{pinatubo,ambit,gupta2018felix,zabihi2018memory,zabihi2019using}},
are particularly suitable for BNN acceleration. BNN computation consists largely of bitwise binary operations, but support is also necessary
for non-binary operations such as pop-count and thresholding. Implementing these by a sequence of gate evaluations is always possible, subject to the limitations of the underlying PIM
technology. 

BNN configuration (filters, weights, thresholds) does not change during inference.
Typical (non-PIM) accelerators fall short of exploiting this, however, due to on-chip 
memory being usually too small to keep all configuration parameters. Thus, repeated data transfer to/from off-chip memory becomes inevitable.
Under PIM, a sufficiently large memory array, on the other hand, renders such data transfers unnecessary. Using PIM for BNN acceleration circumvents expensive data transfers to/from memory. 

In this paper, we introduce \arch, a PIM-based, reconfigurable BNN accelerator for forward propagation. {\arch\ does not use \textit{any} external logic circuitry or sense amplifiers to perform computation. Therefore data never has to leave the memory array.}  \arch\ relies on
a non-volatile (spintronic) PIM technology~\cite{cram}, which provides a better
energy efficiency and storage density trade-off when compared to volatile
alternatives.  The basic spintronic PIM design from~\cite{cram} cannot support BNN operations.
A key contribution of our paper is introduction of novel memory cell architectures augmented with compute
capability, which in turn give rise to novel array architectures for BNN processing.  
As opposed to~\cite{cram}, \arch\ thereby enables computation to occur on only specified rows, while featuring a lower transistor count per cell.

\arch\
can serve as a standard  spintronic memory (STT-MRAM) array and a computational substrate
simultaneously, and is capable of massively parallel and energy efficient
computation.  We implement MNIST, CIFAR-10, and ImageNet classifiers along with a
genomics kernel for similarity matching in \arch\ arrays and use representative FPGA-based
BNN accelerators as baselines for comparison,  which can achieve
significantly higher throughput and energy efficiency
than competing CPU and GPU based implementations.

In the following, 
Section~\ref{sec:back} covers the basics;
Section~\ref{sec:spinbnn}, the proposed \arch\ design; 
Sections~\ref{sec:setup} and~\ref{sec:eval}, the evaluation;
Section~\ref{sec:rel}, the related work; 
and
Section~\ref{sec:conc}, a summary of our findings.

\section{Basics}
\label{sec:back}
%BNN
\subsection{Binary Neural Networks (BNN)}
\label{sec:nn}
Neural Networks (NN), both fully-connected and convolutional, can serve as 
classifiers. Fully-connected NNs consist entirely of fully-connected
layers. Each fully-connected layer is one-dimensional, i.e., the neurons are
arranged in a single line. Every neuron in each layer is connected by a weight
to every neuron of the preceeding layer. Thus, the input for each neuron is the
entire preceding layer. 
Say $N_{(i,l-1)}$ is neuron $i$ in layer $l-1$. Then, $w_{(i,l-1)-(j,l)}$ is the weight from neuron $i$ in layer $l-1$ to neuron $j$ in layer
$l$. If there are $n$ neurons in layer $l-1$, the weighted sum $s$ for
neuron $j$ becomes: $s = \sum_{i=1}^{n} N_{(i,l-1)} \times w_{(i,l-1)-(j,l)}$. The final
value of $N_{(j,l)}$ in this case is a non-linear function $f$ of 
$s$ under the bias $B_l$:
$j = f( B_{l} + s )$. Common choices for $f$ are
sigmoid, tanh, htanh,
ReLu, and sign functions.
To compute the
entire layer $l$, this procedure gets repeated as many times as their are neurons in layer $l$. Once the entire
layer is computed, it is used as input for the proceeding layer.  

Convolutional
networks contain convolutional and pooling layers, in addition to
fully-connected layers. The input and output of convolutional and pooling layers
are three-dimensional collections of neurons, called feature maps (fmaps).
Each neuron in an fmap has 3 coordinates -- (x, y, z) corresponding to (width,
height, depth), respectively -- 
rather than a simple index, as in fully-connected layers. It is this spatial
arrangement of neurons that enables convolutional networks to detect patterns.

\begin{figure}[tp]
\includegraphics[scale=0.45]{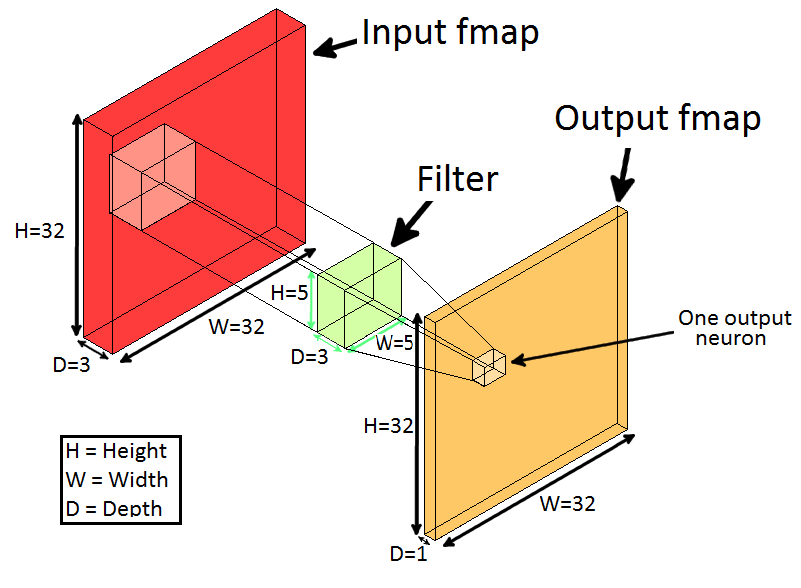}
\caption{Application of a 5$\times$5$\times$3 filter. 75 neurons in the input
fmap are multiplied with the 75 weights in the filter, producing 1 neuron in the
output fmap. Application of the filter to every position generates one layer of
the output fmap. Application of additional filters generates additional layers in
the output fmap.}
\label{fig:conv_layer}
%\vspace{-0.5cm}
\end{figure}

Convolutional layers use filters, which simply are three-dimensional collections
of weights. The height and width of filters can vary in size ($3
\times 3$ to $11 \times 11$, e.g.), but filters typically contain weights to every
layer of the input fmap. For example, if the input fmap has a depth of 3 (such
as an RGB image) and the applied filter is $5 \times 5$, the filter will contain a collection of $5
\times 5 \times 3$ (75) weights.
Applying the filter at one position produces one neuron of the output fmap. The
application of the filter is identical to the calculation of $s$ (and subsequently $f$) 
in a fully-connected layer, except that the weights cover only a subset of the
input neurons in this case. 

To be more specific, 
the filter is
positioned on top of the input fmap, overlapping a subset of the neurons, where
each weight multiplies the neuron it overlaps with. The sum of these products, $s$, is then used
as input to some non-linear function, in the same way as in fully-connected networks.
Fig.\ref{fig:conv_layer} shows an example. Producing one layer of the output fmap entails 
repeating this procedure (by moving the filter
over the input fmap) to generate an output neuron at each position. 
Using multiple filters in a similar fashion produces multiple
layers of the output feature map. 

The stride -- the distance the filter moves by to calculate the next neuron in the output
fmap -- and whether the filter is permitted to slide over the edges of the input
fmap, determine the height and width of the output fmap. If the stride is equal
to 1, and the filter is allowed to slide over the edges of the input fmap, the
output fmap becomes of the same height and width as the input fmap. To allow the filter
to slide over the edges, typically the input is 0-padded, i.e., any weights that
lie beyond the input fmap boundaries are multiplied by 0. 

Pooling layers down-sample fmaps. In a
pooling layer, the fmap is divided by width and height into multiple
sections. A typical pool size would be $2\times 2$. The largest value in each section (in each
layer) is kept and the rest discarded. Thus, each section reduces
to a single neuron. A pool size of $K \times K$ decreases the height and width by a
factor of $K$. Since pooling applies to all depths of the fmap, the
depth remains unchanged.

Reducing the bit precision is a common technique to improve NN efficiency. At the extreme lies 
binarization, where representations for all neurons and weights reduce to only one bit~\cite{BINARIZED}. 
This greatly simplifies hardware complexity. In standard NNs, most of the
operations comprise high latency and power hungry multiply-and-accumulate (MAC), where the values of neurons are multiplied with the values of
weights, which are then summed and transformed non-linearly (typically via
application of sigmoid function). For binary NNs, bit-wise XNOR operations
replace multiplications. The resulting bits are then summed and compared to
a threshold value. Due to their simpler nature, these operations all can be
performed much more quickly and energy efficiently. Note that this applies only
to forward propagation. Inference consists entirely of forward propagation and
thus can fully exploit these benefits. When training, additional non-binary
parameters must be maintained and updated.
We focus on inference in this work and assume that training is
performed offline in software. That said, training can also benefit from more efficient forward propagation as enabled by \arch.

%%%%%%%%%%%%%%%%%%%%%%%%%%%%%%%%%%%%%%%%%%%%%%%%%%%%%%%%%%%%%%%%%%%%%%%%%%%%%%%%%
%%%%%%%%%%%%%%%%%%%%%%%%%%%%%%%%%%%%%%%%%%%%%%%%%%%%%%%%%%%%%%%%%%%%%%%%%%%%%%%%%

%

%%%%%%%%%%%%%%%%%%%%%%%%%%%%%%%%%%%%%%%%%%%%%%%%%%%%%%%%%%%%%%%%%%%%%%%%%%%%%%%%%
%%%%%%%%%%%%%%%%%%%%%%%%%%%%%%%%%%%%%%%%%%%%%%%%%%%%%%%%%%%%%%%%%%%%%%%%%%%%%%%%%

% CRAM
\subsection{Spintronic Processing In Memory (PIM)}
\label{sec:cram}
\noindent Without loss of generality,
\arch\ uses  
Computational RAM (CRAM) as the spintronic PIM substrate~\cite{cram}.  The structure consists of an array of magnetic tunnel
junctions (MTJs) and can be used as a standard STT-MRAM memory array. 
The similarity of \arch\ to MRAM gives it some advantages over other PIM technologies.
%\b{The similarity of \arch\ to MRAM gives it some advantages over other PIM technologies. Notably, the high ON/OFF ratio of the CMOS access transistors prevents large leakage currents and the circuitry of individual cells are well isolated. Hence, we assume that we can perform operations in the array free of error. This is in contrast to crossbar type architectures, such as Intel's x-point, where corruption of neighboring cells is possible due to voltage spikes. } 
Due to
desirable properties of MTJs, the memory is fast, low power, high density, and
non-volatile. Due to non-volatility, standby power is near zero. At the same time, MTJs are inherently compatible with logic operations. This
enables computation to take place {\em entirely} inside the memory array, without the
use of external logic or sense-amplifiers. This provides true PIM capability.
Further, the structure of the array allows for massively parallel computation.
  As we will detail in Section~\ref{sec:spinbnn}, however, the basic array structure from~\cite{cram} cannot support BNNs. 
Building upon this limited basic design~\cite{cram}, a key contribution
of our work is novel cell and array architectures enabling efficient BNNs in memory.

{The magnetic tunnel junction (MTJ) is the key
building block of the memory cells. MTJs
are resistive memory devices, characterized by two distinct resistance levels
to represent logic 1 and logic 0.  Each MTJ has two magnetic layers, a fixed layer
and a free layer. The polarity of the free layer can change but the fixed
cannot. When the two layers are aligned, the MTJ is in the parallel (P) state,
which is considered to be logic 0. If the magnetic layers are not aligned, the
MTJ is in the anti-parallel (AP) state, which is considered to be logic 1. The
MTJ has a much higher resistance in the AP state than it does in the P state. 
The state can be changed by passing current through the MTJ, where the direction }
determines the final state. We now describe the 2T1M design from \cite{cram}.

\begin{wrapfigure}{l}{0.25\textwidth}
%%%%%\vspace{-.3cm}
\includegraphics[width=.97\linewidth]{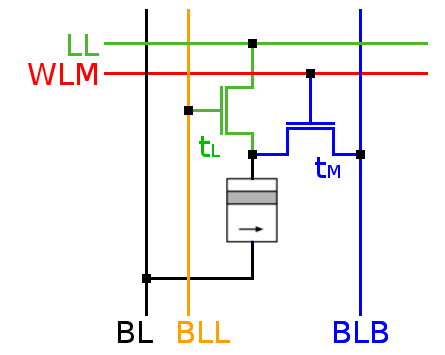}
\caption{2T1M cell~\cite{cram}.}
\label{fig:2Tcell}
%%%%%\vspace{-.3cm}
\end{wrapfigure}

Fig.\ref{fig:2Tcell} shows the default 2T1M cell architecture from~\cite{cram}, which
we include here as a baseline for comparison.
The array formed by these cells is identical to a standard 1T1M STT-MRAM array, except for an 
additional access transistor per cell; the control signal {\em Logic Line}
(LL) which runs along the rows; and the control signal {\em Bit Line for Logic}
(BLL) which runs along the columns.

If no computation is taking place, the array based on the cell from Fig.\ref{fig:2Tcell} can serve exactly as an STT-MRAM array.
In this case, the WLM signal activates the access transistor $t_{M}$ (transistor for
memory), that connects MTJs to the bitline bar (BLB) on a per row basis. Cells
can then be read or written via bitline BL and BLB. The extra hardware enables
computation within the array. The second access transistor $t_{L}$
(transistor for logic), connects the MTJs to the LL. 
BLL controls this second transistor $t_{L}$.
When BLL is activated for multiple columns,
the MTJs 
in those columns, that are also in the same row, are connected to each other
via the shared LL. 
MTJs that are connected over LL in a row can act either as inputs or outputs of
a logic gate. Hence, the result of the logic operation changes the state of the MTJ designated as the output {\em in place}. Voltages applied to BL and BLL control the type of the logic operation. In other words, BL and BLL voltages 
can enforce a specific switching activity (on the output MTJ), which evolves as a function of the resistances (i.e., logic states) of the input MTJs. These voltages thereby can trigger the switching of the output MTJ according to a specific truth table.  A distinct voltage range characterizes each truth table (logic gate). This structure can support different universal sets of Boolean gates and gates of various numbers of inputs~\cite{cram}.

\section{\arch: BNN in Spintronic NVM}
\label{sec:spinbnn}
The 2T1M based design~\cite{cram} from Section~\ref{sec:cram} is limited, as any logic gate activated  in a row also gets activated  in {\em all} rows in the array. This is because BLL (which activates cells to serve as logic gate inputs or outputs) runs through all columns in the array.
If BLL is set for a column, MTJs in that column, in all rows, get connected to LL. While such massive row level parallelism may be desirable, it impairs direct adaptation of the 2T1M design for BNN processing.

\subsection{Spintronic PIM Substrate}
\label{sec:spinCell}

To enable BNN, we consider two different cell (array) configurations, which range from
1T1M (one transistor per MTJ or magnet) to 3T1M (three transistors per MTJ or magnet).

\begin{wrapfigure}{r}{0.25\textwidth}
%\vspace{-.3cm}
\begin{minipage}{\linewidth}
\centering
\captionsetup[subfigure]{justification=justified, width=\linewidth}
    \includegraphics[width=\linewidth]{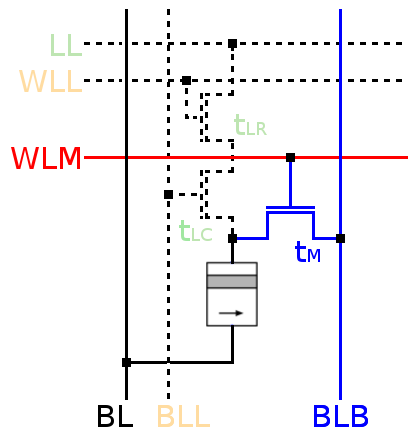}
    \subcaption{3T1M (memory)}
    \label{fig:cell_memory}
\par\vfill
    \includegraphics[width=\linewidth]{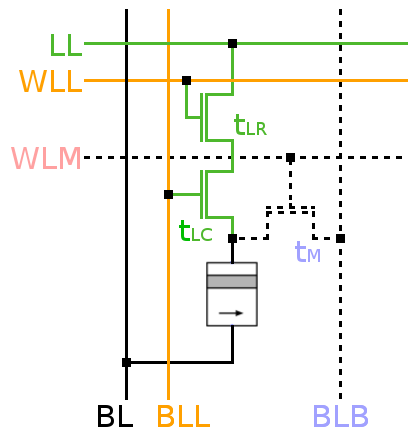}
    \subcaption{3T1M (logic)}
    \label{fig:cell_logic}
\end{minipage}
\caption{3T1M cell.}
\label{fig:cell}
%\vspace{-.3cm}
\end{wrapfigure}

We first describe the 3T1M design.
For efficient BNN processing, we need to perform
computation in only a subset of the rows, while leaving all other rows
unperturbed, which is not the case for the 2T1M design~\cite{cram}. 
To achieve this, we can add a 3rd access transistor, as shown in
Fig.\ref{fig:cell_memory} and \ref{fig:cell_logic}. This design preserves the very
same memory interface as a standard STT-MRAM otherwise.

For logic operations, $t_{L}$ from Fig.~\ref{fig:2Tcell} now becomes
$t_{LC}$ (transistor for logic column). BLL controls $t_{LC}$.
The new signal {\em Wordline for Logic} (WLL) controls the third transistor, $t_{LR}$ (transistor for logic
row).
$t_{LR}$ is in
series with $t_{LC}$, thus, both $t_{LC}$ and $t_{LR}$ must be activated to
connect the MTJ to LL. This enables a row-wise and column-wise specification for
any MTJ to serve as an input or output to a logic gate. BLL and WLL determine
this specification.
As all MTJs in a row share
the same LL, still, only one logic operation can be performed in each row at a
time. However, each operation can be performed in any number of the rows
simultaneously. Hence, the array still features row level parallelism. 
We next take a closer look into memory and logic semantics.

Fig.\ref{fig:cell_memory} shows a 3T1M \arch\ cell
activated for memory. 
WLM is set for only one row. This activates
$t_{M}$ and connects MTJs to the BLBs. MTJs can then be read or written via
voltages applied to BL and BLB. In this configuration, the array acts exactly as a
standard STT-MRAM array. For data retention, i.e., if no read or write access is
the case, keeping WLM, WLL, and BLL at logic 0 suffices.

Fig.\ref{fig:cell_logic} shows a 3T1M \arch\ cell activated for logic.
WLL is set for all rows in which computation should occur. 
In this case we activate rows sequentially (following a similar method to Pinatubo~\cite{pinatubo}). 
This
activates $t_{LR}$ for all cells in the row. BLL is set for all columns that
contain the inputs and the output. This activates $t_{LC}$ for all cells in the
corresponding columns. Since $t_{LR}$ and $t_{LC}$ are in series, only MTJs that
are in cells with  both activated get connected to LL. 
Voltages are applied to BL and BLL then to specify the type of the logic operation.

\begin{wrapfigure}{l}{0.22\textwidth}
%\vspace{-.3cm}
\includegraphics[width=.87\linewidth]{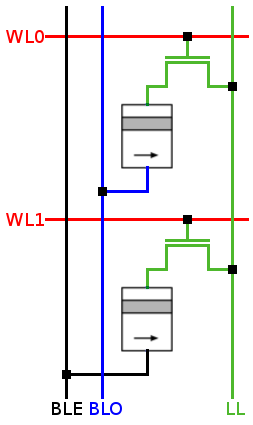}
\caption{1T1M cell.}
\label{fig:stt}
%\vspace{-.3cm}
\end{wrapfigure}

We now describe the 1T1M design. 
While the 3T1M design enables selective computing in the \arch\ array (which is
not the case for the 2T1M design~\cite{cram}), it incurs the area overhead of an extra
transistor.
We next present a more area efficient alternative.
By ``rotating'' the alignment of logic operations relative to the
direction of memory operations, we can reduce the number of transistors from
3 to 1, as shown in Fig.\ref{fig:stt}. 
Due to the ``rotation'' we refer to this design as \textit{transposed}.
The two MTJs from Fig.\ref{fig:stt} are in adjacent cells in the same column. BL and BLB are replaced
with {\em Bitline Even} (BLE) and {\em Bitline Odd} (BLO). Adjacent MTJs are
connected to different bitlines in this case, BLE and BLO, respectively -- BLO and BLE alternate throughout the entire column.  {\em Logic Line} (LL) serves both
as the connection between MTJs for logic operations, and also as input for read
and write operations. {\em Wordline} (WL) controls the single access transistor
in each cell, and is used to connect the MTJ to LL.

When no computation takes place, performing memory operations entails first setting
WL. Specifically, WL gets activated for only one row. This
activates the access transistors and
connects MTJs (in the respective row) to LL. Voltages can then be applied to LL and either
BLE or BLO (depending on the parity of the respective row) for read and write
operations. When compared to the baseline 2T1M design from Fig.\ref{fig:2Tcell},
for memory access,
LL and either of BLE/BLO can be thought of being equivalent to BLB and BL,
respectively. 

For logic on the other hand, WL is selectively set in multiple rows which connects multiple MTJs in a column to LL.
This creates a connection between all activated MTJs that are in the same
column. Voltages are applied to both BLE and BLO in the columns 
where logic gate inputs and outputs reside. 
These voltages determine the type of logic gates performed, following the same basic principle as 2T1M~\cite{cram} and 3T1M designs.

This design has the extra restriction that inputs must be in rows that are all
connected to the same type of bitline (i.e., either all to BLE or all to BLO exclusively) with the
output connected to the other type: If all inputs are connected to BLE (BLO),
the output must be connected to BLO (BLE). This is because, following the basic logic operation principle from~\cite{cram}, all input MTJs must
be in parallel with each other and in series with the output MTJ, when it comes to forming logic gates -- i.e., to enforcing switching on the output MTJ as a function of the logic states (resistances) of the input MTJs.
This may impact data layout, depending on the algorithm being mapped to the
array.  However, the effect is negligible in most operations, including the entire implementation of BNNs. Due to the reduction to one transistor, the resulting array has
nearly the same density as a standard STT-MRAM array. {This is significant as density is a key advantage of STT-MRAM. }

%%%%%%%%%%%%%%%%%%%%%%%%%%%%%%%%%%%%%%%%%%%%%%%%%%%%%%%%%%%%%%%%%%%%%%%%%%%%%%%%%%%%%%%%%
%%%%%%%%%%%%%%%%%%%%%%%%%%%%%%%%%%%%%%%%%%%%%%%%%%%%%%%%%%%%%%%%%%%%%%%%%%%%%%%%%%%%%%%%%

%%%%%%%%%%%%%%%%%%%%%%%%%%%%%%%%%%%%%%%%%%%%%%%%%%%%%%%%%%%%%%%%%%%%%%%%%%%%%%%%%%%%%%%%%
%%%%%%%%%%%%%%%%%%%%%%%%%%%%%%%%%%%%%%%%%%%%%%%%%%%%%%%%%%%%%%%%%%%%%%%%%%%%%%%%%%%%%%%%%

We will next look into basic computational BNN building blocks and 
data layout in the \arch\ array along with design optimizations. Without loss of
generality, in the following, we will use the 3T1M \arch\ as a running example. Transposed 1T1M \arch\ simply
rotates the sense of logic operations to occur in columns rather than in rows,
and is logically equivalent to 3T1M.

%\vspace{-0.4cm}
\subsection{Computational Building Blocks}
\label{sec:basiccramoperations}

\noindent 
By construction, \arch\ can perform any logic gates that CRAM can perform~\cite{cram}, 
as the principle for gate formation and logic operation is the same in spite of
the differences in cell and array architectures. 
This translates into a universal set of
gates including NOT, NAND, NOR, AND, OR, and MAJ(ority). 
The number of inputs is arbitrary but
limited by voltage variation
(Sect.\ref{sec:noisemargin}). 
The operating principle is as follows, irrespective of the cell type: First, all MTJs
(or magnets) to serve as either input or output of the logic gate being formed are
connected to each other. 
%to {\em Logic Line} LL. 
The logic states correspond to the resistance
levels of participating magnets. This always
renders the same topology of a resistive network: all input magnets in parallel,
connected in series to the output magnet. Voltage $V$ applied on the respective
bitlines, to set the gate type $G$, forms a voltage differential across this
resistive network.  The magnet corresponding to the gate output 
is preset to a known value.  $V$ forces the gate output to switch or preserve
its value as a function of the state of the inputs, according to the truth table
of $G$. More specifically, $V$ (along with the state, i.e., resistance levels of
the input magnets) determines the current through the output magnet, which
results in switching if this current exceeds the switching threshold.

\begin{wrapfigure}{l}{0.55\textwidth}
%\vspace{-0.3cm}
\includegraphics[width=\linewidth]{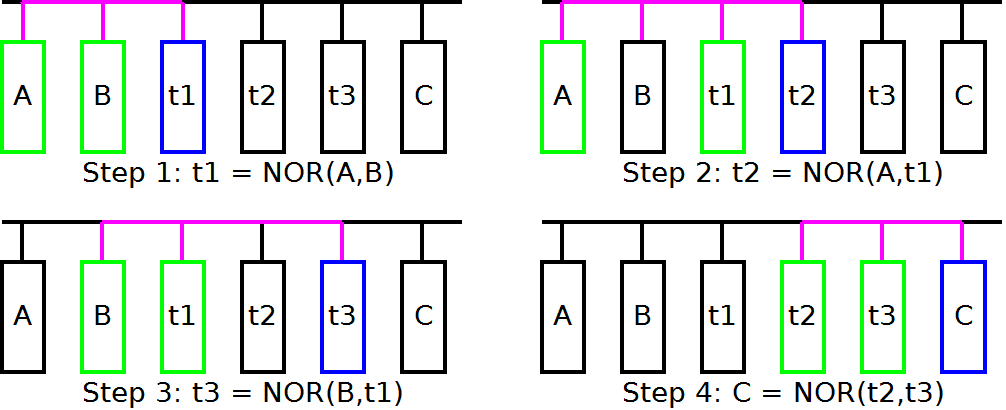}

\caption{XNOR (inputs in green, outputs in blue).}
\label{fig:xnor}
%\vspace{-0.3cm}
\end{wrapfigure}

XNOR is a critical component of binarized NN implementations. The
output of XNOR is 1 when the inputs are the same, and 0 otherwise. 
\arch\ cannot support XNOR operation in a single step (following the resistive divider based principle).
However, \arch\ can perform XNOR using a sequence of NOR
gates, following:
$A \otimes B = ((A + (A + B)' )' + (B + (A + B)' )' )'$.
This process requires four NOR gates and thus four steps to complete, in
addition to
three temporary values (t1, t2, t3, respectively), as shown in Fig.\ref{fig:xnor}. The dual implementation is also possible, using two NOT and three NAND gates.

Addition is implemented by
the ripple-carry algorithm, without loss of generality, where the output is
computed one bit at a time. The first step is a half add and all remaining steps
are full adds, which consist of NAND
and NOT gates. Each full add
requires four temporary values, including the carry bit, and takes a total of 5
steps. Thus, addition of two n-bit numbers requires 5n steps and 4n temporary
values. {Using only NAND gates, addition takes 9n steps.}

Comparison to a threshold value 
%and subsequent ``sign'' operation 
is a common BNN correspondent of the non-linear function $f$ from
Section~\ref{sec:nn} for ordinary NN.
We implement this by subtracting the threshold value from the input and then
taking the sign of the result. The sign is equal to the inverse of the borrow
out signal when subtracting the threshold from the input. Thus, we do not need
to perform the actual
subtraction, and we only need to compute the borrow
out signal. 
To this end, 
we use the ripple-borrow algorithm where a full-subtractor
is implemented for each bit of the input. However, for each bit, we do not
compute the difference bit, but only the output borrow, $B_{out}$. Given $x$, a bit
of the threshold value, and $y$, a bit of the input, the equation for $B_{out}$
in NAND form is as follows:
$B _{out} = ((x'B_{in})'(x'y)'(yB_{in})')'$,
where $B_{in}$ is the input borrow signal, which is set to 0 for the least
significant bits. The $B_{out}$ signal generated is used as the $B_{in}$ for the
next full-subtractor. This is repeated for all bits of the input. The final
$B_{out}$ bit is then inverted to produce the sign bit. Computing $B_{out}$ for
each bit requires one NOT gate, 4 NAND gates, and 4 temporary values. Thus,
comparing two n-bit numbers takes 5n+1 steps and 5n temporary bits.

Popcount takes a list of bits as input, and produces an integer equal to the number of 1s in the list. \arch\ implements popcount with a sequence of adds. Additions are
scheduled in a hierarchical manner, following the scheme of a full adder tree.
Initially there is a list of 1-bit operands. First, all bits (operands) are
paired and added together to form a list of 2-bit numbers. Then, pairs of these 2-bit numbers are added to form 3-bit numbers. This
process repeats until there is only one number remaining, which has a value equal to the
popcount of the original list of bits. If at any point the number of operands is odd, the one extra operand is sign extended and carried over to the next step. Affine transformation enables bits that are 0 to act as -1 \cite{FPBNN}, to align with the original algorithm \cite{BINARIZED}. This is done by doubling (logical left shifting) the final popcount result and subtracting from it a number equal to the number of input bits. In \arch, the doubling operation can be done automatically by storing the result pre-shifted and the subtraction can be embedded in the subsequent thresholding operation. Thus, the transformation creates no additional overhead.

Batch Normalization primarily serves accelerating
training but can also improve accuracy during inference.
Batch normalization consists of a
scale (multiply) and  a shift (add) transformation. In the context of BNNs, it
is applied to the result of the popcount before thresholding occurs.
Shift-Based batch normalization (which simply replaces the
costly multiplication with a shift) incurs a negligible
accuracy loss~\cite{BINARIZED, FPBNN} and 
works particularly well
in \arch: The shift operation consists simply of writing 0s in the appropriate
locations, which overwrite either the most or least significant bits of the
input. No data transfer is necessary. Then, addition proceeds
as described above.
Other networks~\cite{FINN} 
implement the same
effect by 
thresholding and a modification of the input weights. { XNOR-Net \cite{xnornet}, on the other hand, uses standard batch normalization.}

\subsection{BNN Data Layout}
\begin{figure}[ht]

\centering
\includegraphics[scale=0.17]{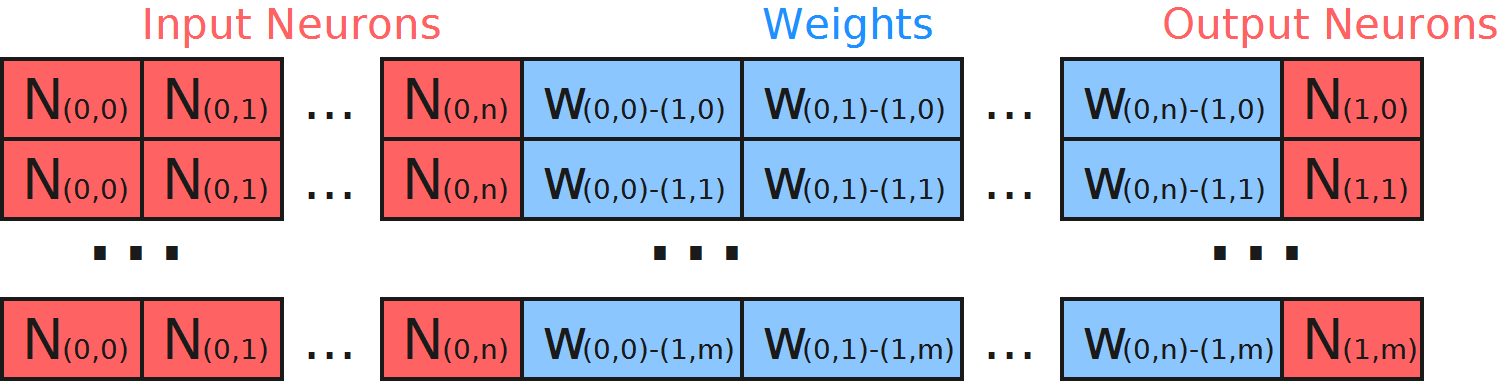}

\caption{Layout for fully-connected forward propagation. N$_{(i,j)}$:
neuron j in layer i. $w(i,j)$-$(k,l)$: weight from neuron j in layer i to neuron l in layer  k, where k = i+1. i=0 and k=1 in this example.}
\label{fig:fullyconnected}
%\vspace{-.3cm}
\end{figure}

In fully-connected forward propagation, each neuron has as inputs all neurons in
the previous layer. Hence, for each neuron in the layer, there is a set of weights to
every neuron in the previous layer. 
\arch\
uses 
data duplication to compute each neuron 
in the output layer in parallel, as follows:
We group one or more rows together and dedicate {them} to computing
one neuron in the output layer. Such grouping exploits row-level parallelism at a finer granularity, i.e., in computing each output neuron. Each group of rows 
%must 
contains a copy of every
neuron in the previous layer and all weights to those neurons. For example, when computing
forward propagation from a layer with $n$ neurons to a layer with $m$ neurons,
there are $m$ groups of rows, each of which contain $n$ neurons and $n$ weights.
Each group also stores  any necessary bits for batch normalization and
thresholding, and computes in parallel.

The optimal group size $g$, i.e., the number of rows in each group, depends on
the network layer sizes.
Larger $g$ increases
parallelism but introduces extra overhead due to communication between rows.  
{This is because the results from each row within a group must be moved to a single row in order to compute the final result. The number of extra data transfers required increases with group size.}

Fig.\ref{fig:fullyconnected} provides an example layout where each group is a
single row ($g=1$). The input is layer 0; the output, layer 1. Each of the $n$ input
neurons are duplicated on $m$ rows. The neurons are then XNORed with the
corresponding weights, where
the results overwrite the input neurons (as a space optimization). These
bits are summed next (popcount),
modified by batch normalization,
and finally thresholded.
%as described in Section \ref{sec:basiccramoperations}.
All neurons of the computed
layer are then stored in a single column (as demarcated by {\em Output Neurons} in
Fig.\ref{fig:fullyconnected}). The output neurons 
%must then be read out,
%which 
can be either the final output (if corresponding to the last layer), or inputs to the next layer of
the network.

In 3-dimensional convolution, a filter is a 3-dimensional collection of weights.
Filters can have different heights (y dimension) and widths (x dimension), but
all filters have weights to all layers (which comprises the depth, z dimension) of the input fmap
(Sect.\ref{sec:nn}).
In other words, the depth of the filter must be the same as the number of layers
(depth) of the input fmap. 
Each filter  generates one layer of the output
fmap. Thus, 
there are as many layers in the output fmap as there are filters.
%of the input fmap. 

To compute the value of an output neuron, each weight (a single
bit for BNN) of a filter is XNORed with a neuron in the input
(also a single bit). The number of 1's resulting from these XNOR operations are
then counted. This sum is next batch normalized and thresholded, to render
the bit value of the output neuron. The filter is then slid (in the x and y
dimensions) over the input and an output neuron thereby is computed at each position. 

\begin{figure}[ht]

\includegraphics[scale=0.15]{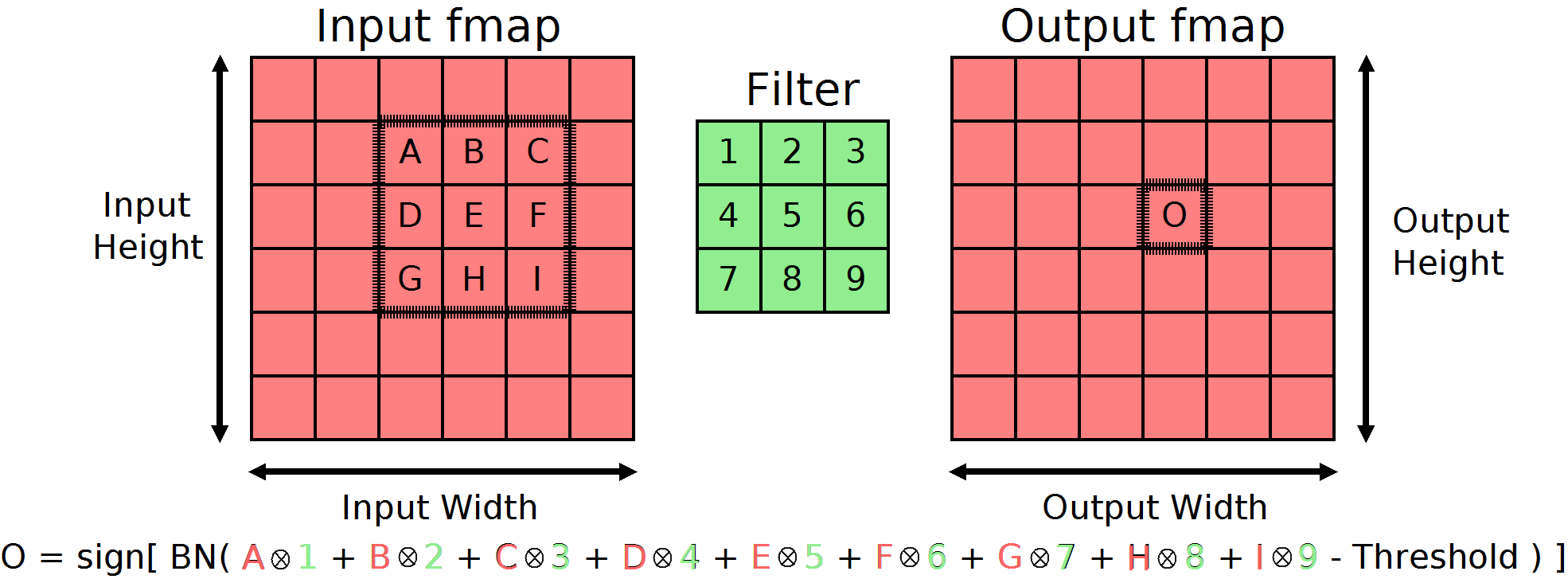}

\caption{Convolution in 2 dimensions. Explicitly marked are the filter position in the input fmap and the location of the (output) neuron being calculated in the output fmap. A, B, ..., I, O indicate neuron values; 1, 2, ..., 9, filter bits. BN: Batch Normalization.}
\label{fig:convolution}
%\vspace{-.3cm}
\end{figure}

\begin{figure}[ht]

\includegraphics[scale=0.15]{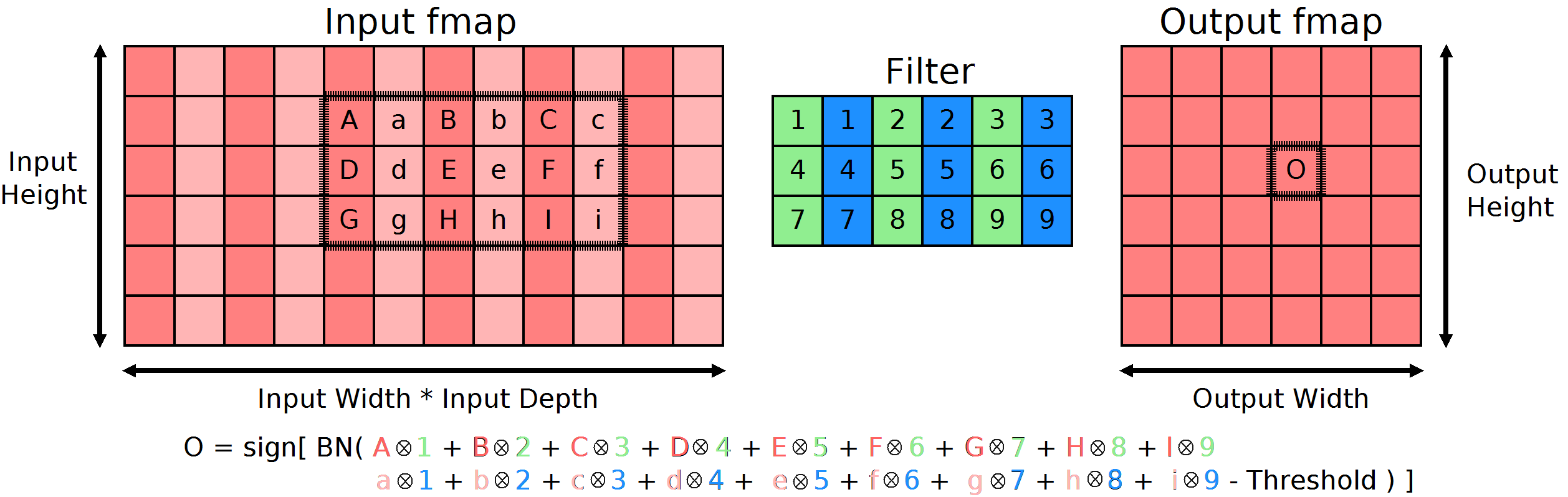}

\caption{Convolution in 3 dimensions. Explicitly marked are the filter position in the input fmap and the location of the (output) neuron being calculated in the output fmap. A, B, ..., I, O indicate neuron values; 1, 2, ..., 9, filter bits. Lower case neurons and weight bits shaded in blue are at depth 2. BN: Batch Normalization.}
\label{fig:convolution3d}
%\vspace{-.3cm}
\end{figure}

Fig.\ref{fig:convolution} depicts 
an example, where the z dimension (depth) is 1. To compute the output neuron labeled O, each filter
bit is XNORed with the input neuron it is overlapping (filter bit 1 with neuron A,
filter bit 2 with neuron B, ...). 
%The filter would then be shifted to the
%right and the output neuron to the right would be computed accordingly.
Fig.\ref{fig:convolution3d} provides an example of
3-dimensional convolution with depth=2.
In this case, neurons and weights at consecutive depths are stored in consecutive cells in the
\arch\ array.

Output neurons can be computed in parallel as they are not data dependent.
Computing output neurons in parallel in \arch\ requires data duplication. As is
done for fully-connected layers, rows are grouped together and each group
computes one output neuron. Any input neurons needed for the computation are
written into the same group of rows, along with bits for the filter, batch
normalization, and thresholding. Thus, each group contains all required data and
can operate independently from all other groups. 

\begin{wrapfigure}{l}{0.30\textwidth}
%\vspace{-.3cm}
\includegraphics[width=.97\linewidth]{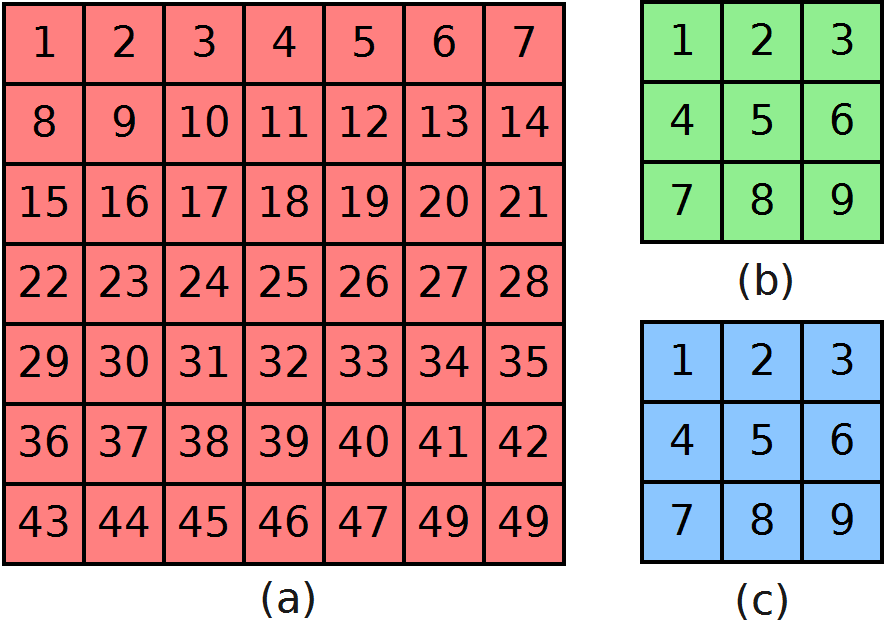}

\caption{Logical arrangement of input neurons (a); of two filters (b), (c).}
\label{fig:convInp}
%\vspace{-.3cm}
\end{wrapfigure}

\begin{figure}[ht]

\centering
\includegraphics[scale=0.10]{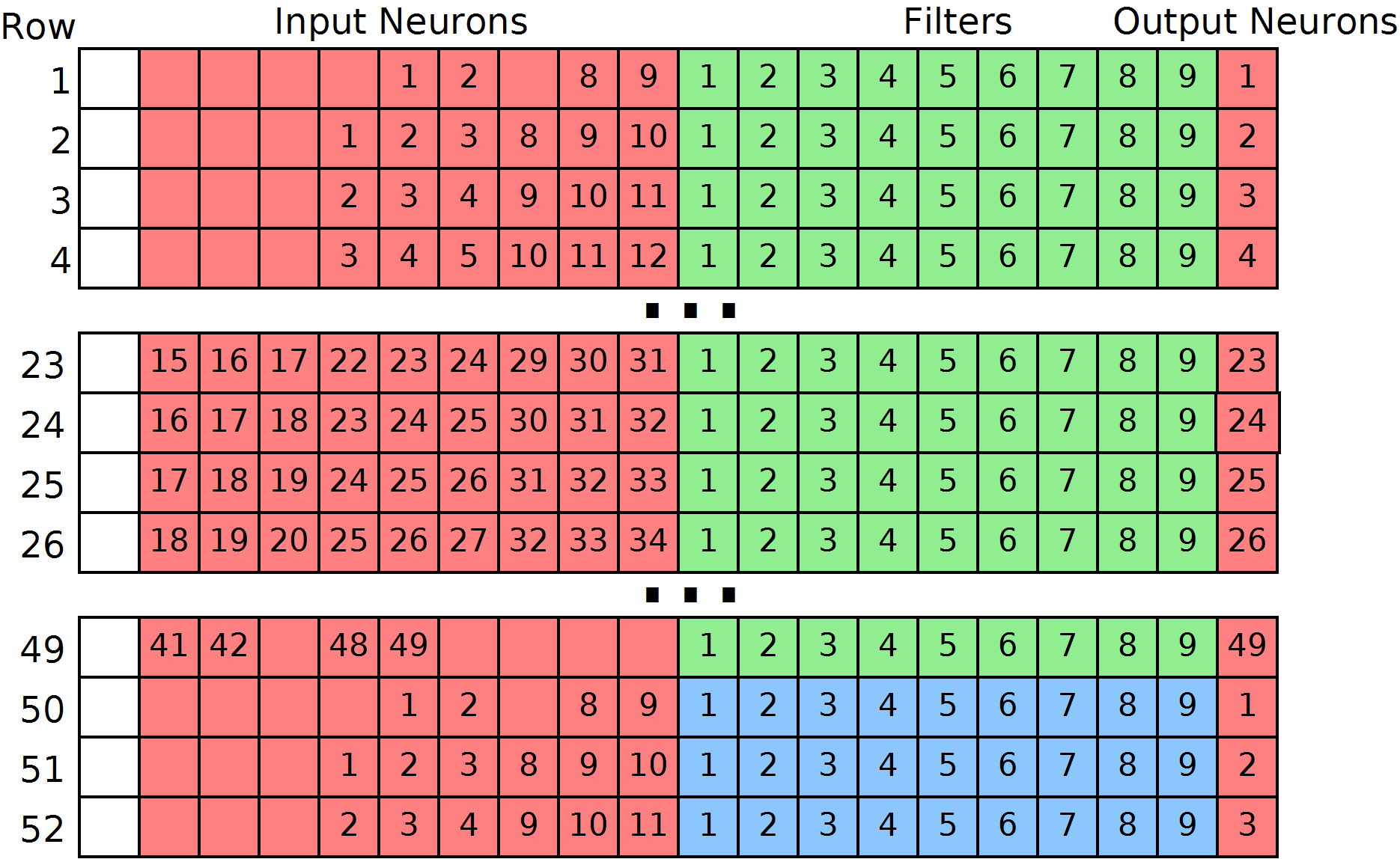}

\caption{The spatial arrangement of input, filter, and output bits during
convolution in \arch. Bits of neurons and filters are labelled according to
their logical position in Fig.\ref{fig:convInp}. All
filters and all locations of each filter are computed in parallel, exploiting \arch's row-level parallelism. For clarity,
bits for batch normalization and the threshold are not shown.}
\label{fig:convDup}
%\vspace{-.4cm}
\end{figure}

Fig.\ref{fig:convInp} shows an example 2-dimensional
(7$\times$7$\times$1) input layer along with two 2-dimensional (3$\times$3)
filters. Fig.\ref{fig:convDup} captures the placement of these bits in
the \arch\ array, for a group size $g$ of 1 (i.e., each row computes one output
neuron).

Since the filter is 3$\times$3 and the input is 1 layer deep, there are
3$\times$3$\times$1 = 9
filter bits. All 9 of the bits of each filter are written into a single row (in
the {\em Filter Bits} portion of the \arch\ array from
Fig.\ref{fig:convDup}).
They are then duplicated across the rows so that each output bit being computed
has its own copy. 

In Fig.\ref{fig:convDup}, each row of filter bits on rows 1-49 are all
the same and contain the bits for the first Filter
(Fig.\ref{fig:convInp}(b)). The second filter is duplicated
(Fig.\ref{fig:convInp}(c)) on rows 50-99. If there were
additional filters, they would be written in the same manner on the following
rows. The 9 filter bits can overlap with up to 9 input neurons, and which
neurons these are depends on the specific position of the filter. 

The input bits that the filter overlaps at each position are
written to separate rows. Thus, there is a row for each possible position of the
filter and each position of the filter is computed simultaneously. Since all
filters are applied to the same input neurons, the input bit pattern must be
repeated for each filter. For example, in Fig.\ref{fig:convDup}, row
1 is repeated on row 50 for the computation of the 2nd filter. Row 2 is repeated
on row 51, and so on. 
Note that the filter can go over the edge of the input
fmap. For example, the 3$\times$3 filter centered at input neuron 1 will only overlap
input bits 1, 2, 8, and 9 (as captured by rows 1 and 50). Thus, the remaining 5 locations of the 9 locations
allocated for the input bits are left empty. These are filled with dummy 0's.

As computation in each row is independent, XNOR, addition, batch normalization,
and thresholding operations can be performed in parallel for all output bits of
every filter. At the end of these steps of computation, all output bits are
stored in a single column. 
The example is 2-dimensional for clarity, however, this layout easily extends
to 3-dimensional convolution. If the input is z layers deep, each input and
filter bit becomes an array of z bits stored in consecutive cells. For k
by k filters, each group of rows contains k$\times$k$\times$z input and filter bits in this case.

%%%%%%%%%%%%%%%%%%%%%%%%%%%%%%%%%%%%%%%%%%%%%%%%%%%%%%%%%%%%%%%%%%%%%%%%%%
%%%%%%%%%%%%%%%%%%%%%%%%%%%%%%%%%%%%%%%%%%%%%%%%%%%%%%%%%%%%%%%%%%%%%%%%%%

%%%%%%%%%%%%%%%%%%%%%%%%%%%%%%%%%%%%%%%%%%%%%%%%%%%%%%%%%%%%%%%%%%%%%%%%%%
%%%%%%%%%%%%%%%%%%%%%%%%%%%%%%%%%%%%%%%%%%%%%%%%%%%%%%%%%%%%%%%%%%%%%%%%%%

\subsection{Computation vs. Communication Trade-off}
\noindent The fully-connected and convolutional layer
organizations we covered so far locate the output neurons in a single column.
For non-transposed
\arch\ arrays
this is a bottleneck as each
bit must be read out sequentially. One solution is {\em sequentializing} computations, by
computing not all, but only a subset of the
output layer neurons in parallel. For fully-connected layers, this translates
into a chosen subset of
the output layer neurons being computed in parallel;  for convolutional layers, a
subset of the filters.
This leads to less data duplication. At the same time,
we can store the results of each sequential step 
in separate columns, increasing
the parallelism at which they can be read out.

\section{Evaluation Setup}
\label{sec:setup}

\subsection{System Configuration}
\label{sec:systemconfig}
%\begin{wraptable}{l}{9cm}
\begin{table}[th]

\caption{MTJ Specification. \label{tab:mtj}}

  \centering
  \resizebox{.6\textwidth}{!}{
\begin{tabular}{l||c||c}
%\hline
{\bf Parameter} & {\bf Modern (M)} & {\bf Future (F)} \\
\hline \hline
{\bf MTJ type} & Interfacial perp & Interfacial perp \\
\hline
{\bf Material system} & \begin{small} CoFeB/MgO/CoFeB \end{small}& \begin{small}CoFeB (SAF)/MgO/CoFeB \end{small}\\
\hline
{\bf MTJ diameter} & 45nm & 10nm \\
\hline
{\bf TMR} & 133\% & 500\% \\
\hline
{\bf Threshold current} $I_{C}$ & \SI{40}{\micro \ampere} \cite{modernmtj}& \SI{3}{\micro \ampere} \footnotemark\\
\hline
{\bf Switching time} $T_{switch}$ & 3ns \cite{modernmtj} & 1ns \\
\hline
{\bf R$_P$} & 3.15K$\Omega$ & 7.34K$\Omega$ \\
\hline
{\bf R$_{AP}$} & 7.34K$\Omega$ & 76.39K$\Omega$ \\
%\hline
\end{tabular}
}

\end{table}
%\end{wraptable}
 \footnotetext{Through decrease of the damping constant of ferromagnetic materials \cite{mtj1,mtj3,mtj4} and adopting a dual-reference layer structure \cite{mtj5,mtj6} switching currents of \SI{1}{\micro \ampere} are predicted to be possible. We assume \SI{3}{\micro \ampere } to be conservative.}

\noindent{\bf Technology Parameters:}
While MTJs manufactured today are capable of being used effectively in \arch\
arrays, 
%their performance is 
they are expected to significantly improve in the coming years. 
We therefore consider both a
%evaluate CRAM performance for both 
modern day 
and a projected future MTJ 
specification, listed in 
Table~\ref{tab:mtj}. 
The threshold current, $I_{C}$, is the current at which the
MTJ has a 50\% chance of switching within the switching time $T_{switch}$. By
setting the write current to 1.5$\times I_{C}$, switching occurs with a
probability of error less than $10^{-5}$.

\noindent{\bf Voltage Signature per Logic Gate:}
\label{sec:noisemargin}
\noindent The switching activity of each logic gate output depends on the current 
though the output MTJ. This current in turn is a function of
the voltage applied on the bitlines and the inputs -- i.e., the states or
resistances of the MTJs
which form the gate inputs. 
The applied voltage directly determines the type of the gate. In other
words, the voltage acts as a {\em signature} for the gate type.  Therefore,
correct operation demands preventing potential voltage variation on the bitlines (due to,
e.g., manufacturing imperfections) from making one gate act as another gate.

\begin{wraptable}{r}{5cm}
%\vspace{-.3cm}
%\begin{table}[ht]
\caption{Voltage signatures (ranges) for \arch\ gates in mV. \label{tab:voltageranges}}
  \centering
  \resizebox{0.35\textwidth}{!}{
\begin{tabular}{  l || c  || c  }
&{\bf Modern (M)}&{\bf Future (F)} \\
\hline\hline
NOT & 336 (168) &  172 (191)  \\
\hline
NAND & 243 (59) &  112 (82) \\
\hline
NOR & 202 (25) &  64 (13.6) \\
\hline
IMAJ-3 & 186 (15.9)  & 61 (11.0)  \\
\hline
IMAJ-5 & 161 (5.7)  & 56 (3.8) \\
\end{tabular}
}
%\end{table}
%\vspace{-.3cm}
\end{wraptable}

The voltage signature of a gate 
is not restricted to a single value, and can assume any value in a
gate-specific range, as
shown in Table~\ref{tab:voltageranges} for \arch\ gates. The values in the
parentheses 
capture the range.
The parallel combination of the input MTJ resistances, in series with the
output MTJ resistance, determine the voltage range. Any voltage in this
range facilitates the switching activity of the output MTJ according to the truth table of the
respective gate, for all input combinations. 
Each of these voltage ranges can be interpreted as a gate-specific voltage
margin: Correct operation is guaranteed, as long as any fluctuation 
renders a 
voltage within the range corresponding to
the respective gate.

%\b{
%We also note that 
The parasitic resistance of the access transistors can also affect the voltage signature and margin of the logic gates. The effect on the signature can be accounted for at design time, i.e., 
%meaning 
the voltage signature can be set depending on the expected resistance of the transistors. This is because the expected resistance is technology dependent and remains constant for each operation. Regardless of the operation, there is always one (two) transistors in between each MTJ and the logic line for the 1T1M (3T1M) design. This enables the setting of the proper voltages on the bitlines to drive the operations. What could be problematic is fluctuations of threshold voltage due to process variation, which could change the transistor resistance and consequently narrow the voltage margin. As the transistors can only add resistance, the only pathological case is if the resistance is sufficiently high to prevent switching. Fortunately, the anticipated transistor resistance, \SI{100}{ \ohm}, is significantly smaller than MTJ resistances listed in Table \ref{tab:resistancevalues}. This makes logic gates relatively immune to transistor parasitic resistance. For example, the parasitic resistance would have to be 700 (15,000) $\Omega$ for modern (future) devices in order to cause incorrect operation of a NAND gate. The large voltage margin seen for two-input gates and the high resistance values of the MTJs lead us to conclude that logic operations are robust to parasitic transistor resistance and process variation.  
%}

\begin{wraptable}{l}{7.1cm}
%\begin{table}[ht]
%\vspace{-.3cm}
\caption{Combined input resistance $(\Omega)$ for 2 inputs.
%all  2-input combinations.
  %of MTJ combinations for a gate with fan-in of 2. 
\label{tab:resistancevalues}}

  \centering
%  \resizebox{0.45\linewidth}{!}{
   \resizebox{0.45\textwidth}{!}{
\begin{tabular}{  r || c  || c   }
%\hline
%& \multicolumn{2}{c||}{\bf Modern (M)} & \multicolumn{2}{c}{\bf Future (F)} \\
%  \cline{2-5}
  %\hline\hline
%Gate & STT & SHE & STT & SHE \\
Input State  & {\bf Modern (M)}  & {\bf Future (F)} \\
%Input State  & STT  & STT \\
\hline\hline
%& STT & SHE & STT & SHE \\
%\hline
R$_{11}$: 2 AP, 0 P & 6820 &  50900  \\
\hline
R$_{01}$=R$_{10}$: 1 AP, 1 P & 5354 & 23590  \\
\hline
R$_{00}$: 0 AP, 2 P & 4725  & 19050 \\
%\hline
\end{tabular}
}

%\caption{Bitline voltage inputs (and ranges) for common gates in mV. \label{tab:voltageranges}}

%\end{table}
%\vspace{-.3cm}
\end{wraptable}

NAND and NOR are 2-input; inverted majority IMAJ-3 and IMAJ-5 are 3- and
5-input gates, respectively.  As Table~\ref{tab:voltageranges} indicates,
generally, gates with
larger fan-in have smaller margin. The reason 
is twofold:
First, more input resistances in parallel have a smaller combined resistance. Thus,
smaller changes in voltage cause a larger change in current. Second, differences in
resistance between different combinations of inputs is smaller.
%, even relative to the combined input resistance.  
This effect is made worse by the fact that
the combined input resistance is in series with the output resistance (which is
always the case, 
independent of the number of inputs). Hence, there is a sharp drop in voltage
margin with increasing number of inputs. 

It is noteworthy that NAND has a larger
margin than NOR. 
Table \ref{tab:resistancevalues} depicts the combined input resistance (for
all possible input combinations) for 2 inputs. Recall that an MTJ in Anti-Parallel (AP)
state incurs a higher resistance than in Parallel (P) state. The resistance
in AP state (R$_{AP}$) corresponds to logic 1; in P state (R$_{P}$) , logic 0.
R$_{00} < $  R$_{01}$ = R$_{10} < $ R$_{11}$ applies, where R$_{00}$ captures
the combined input resistance for the input combination 00; R$_{01}$, for 01,
and so on. 
For a given gate (hence voltage signature), the current through the
output magnet, $I$, assumes its maximum for the lowest value of the combined input
resistance, hence I$_{00} > $  I$_{01}$ = I$_{10} > $ I$_{11}$ applies. Let us
assume a non-transposed \arch\ design, without loss of generality, where
the output magnet is preset to 0 to perform NAND or NOR.
For NAND, the output should switch for all combinations but 11; but for NOR, for
only 00. Hence, it is critical for NAND to differentiate between R$_{11}$ and
R$_{01}$ (= R$_{10}$);
for NOR, between R$_{00}$ and R$_{01}$ (= R$_{10}$).
As Table \ref{tab:resistancevalues} indicates, the difference between the combined 
input resistances for NAND are much larger, which renders a larger voltage
margin for correct operation. 

As Table~\ref{tab:voltageranges} indicates, for current MTJ specifications, only
NAND and NOT gates are practical. The voltage margin for gates with 3 or more
inputs are too small to guarantee correct operation. Predicted future MTJs have
a much lower switching current but a significantly larger TMR, i.e., 
(R$_{AP}$-R$_P$)/R$_P$.
The effect of a larger TMR is making differences in the
combined resistance of different input combinations more pronounced, and thereby
increasing the voltage margins. The effect of a smaller switching current is the
opposite.
For low fan-in gates, the TMR effect is dominant and
future MTJs have a larger margin than modern MTJs, despite the lower current.
For high fan-in gates, the opposite is the case.
NOT and NAND gates have very large voltage margins relative to their 
voltage signatures. 
This generally applies to other gates, as well, considering future MTJs.
Restricting logic operations to only NAND and NOT gates significantly reduces
susceptibility to process variation. Luckily, this is a universal set of gates. 
%\b{
Accordingly in the evaluation, we restrict operations to only NAND, NOT, and COPY.
%}

For the array configuration, we use a tiled architecture, where each tile is a single \arch\ array. We evaluate two different arrays sizes, 128KB, with 1024 rows and columns, and 512KB, with 2048 rows and columns. Arrays can be accessed and used for computation independently.  
%}

 For modern MTJ analysis, we take latency and energy estimates for STT-MRAM array accesses from \cite{dong2008circuit}, scaled to our array capacity and node size. As future MTJs are still a
few years from being ready for production, the exact supporting peripheral
circuitry overhead is unknown. To estimate it, we scale the latency and energy estimates so that the peripheral circuitry maintains the same percentage share of each. In both cases, the peripheral circuitry is responsible for a significant percentage of the latency. This renders a close-to-worst-case analysis, without any optimization specific to \arch.

While the method by which \arch\ performs computation is different,
the array structure is similar to~Pinatubo\cite{pinatubo}.
The demands on our peripheral circuitry are similar to \cite{pinatubo}, as well, though
we do not use sense amplifiers during computation. In the case of the 2T1M design (which we include for completeness), we use an extra transistor.
Given MTJs are lower power devices, the current draw remains relatively modest
even during highly parallel computation. Using projections for future MTJ
devices (Table \ref{tab:mtj}), a \arch\ array performing
computation on every row would still consume far less current than a
DRR3 SDRAM write operation \cite{micron}.

Since \arch\
requires multi-row access, we follow the approach developed in \cite{pinatubo},
where row addresses are supplied sequentially. Modifications to the local
wordline driver allow these to be latched until cleared. For
logic operations, where multiple rows can be activated, we assume that the rows
are addressed sequentially and take into account the latency and energy of both
the row activations and voltages applied to the bitlines.

As for system integration, \arch\ can be attached to the host system as a standalone accelerator over PCIe,
or be part of the memory hierarchy (which renders a similar interface to
emerging hybrid memory systems featuring NVM). In any case, \arch\ memory space
is exposed to the host (e.g., over Direct Memory Access). {Software would utilize \arch\ in a similar manner as a GPU. Code written for \arch\ must be hardware aware in order to utilize its resources efficiently. However, unlike a GPU, \arch\ will likely be capable of storing all required inputs, making data transfers between \arch\ and the disk relatively infrequent. If the entire memory stack is implemented in STT-MRAM, \arch\ could comprise the entire memory, enabling computation anywhere in the memory stack.}

We assume that all BNN configuration parameters including weights and thresholds
are stored in the arrays prior to computation.  
These parameters do not change during inference.
Additionally, the \arch\ arrays have sufficient storage to hold all, along with sufficient space for temporary values. Thus,
there is no need to communicate with the host during inference.
Different layers of the networks are performed in
different collections of \arch\ arrays. This creates an opportunity for pipe-lining. 
Inference constitutes two phases:
a highly parallel computation phase and a data communication
and duplication phase. 
Thus, each layer of the implemented networks has an
associated latency and energy cost for both computation and communication. 
The
computation phase comprises logic and memory operations which move
data within the arrays. The communication phase comprises data transfers
between arrays, which consists of reads from the source array and writes to the
destination array. 

\subsection{Benchmarks and Baselines for Comparison}
\noindent We implement one Imagenet, two MNIST, and two CIFAR-10 classifiers in \arch, and use two representative FPGA-based BNN accelerators
as the baselines for comparison, {\bf FP-BNN}~\cite{FPBNN} and {\bf FINN}~\cite{FINN}, respectively,
which achieve higher throughput and better energy efficiency
than GPU-based implementations.  FP-BNN~\cite{FPBNN} uses a Maxeler MPC-2000 with a Stratix V FPGA, and FINN~\cite{FINN} is implemented on the Xilinx Zynq-7000 SoC ZC706 Evluation kit.
We reproduce the same network topologies in {\arch}~\footnote{Some networks take in non-binary inputs. 
We handle multi-bit precision following~\cite{BINARIZED}, by
computing bits of each significance
separately, in the same manner as binary inputs. 
}.
While the supporting hardware is different,
the inputs, network sizes, and operations performed are logically identical. 
As a result, \arch\ results in the same output and accuracy. In other words, we
perform an iso-accuracy comparison. We also implement BioNET, a BNN kernel for similarity matching in genomics on \arch.
To quantitatively characterize \arch's performance and energy efficiency, we 
use an event-driven
in-house simulator.

Fully-Connected FP-BNN follows the layer configuration of \cite{FPBNN}. The model consists of 4
fully-connected layers. The input is a 28$\times$28 gray scale image with 8-bit pixels.
There are 784 input neurons (each 8 bits), 10 output neurons, and three hidden
layers of 2048 neurons each. The network achieves an 98.24\% accuracy on the
MNIST dataset.

Fully-Connected FINN, the fully-connected network in \cite{FINN}, is slightly smaller. The input is
28$\times$28 images that have been binarized. There are 3 hidden layers, each has 1024
neurons. The output is a vector of 10 bits corresponding to the 10 output
neurons. It achieves a 98.4\% accuracy on the MNIST dataset.

Convolutional FP-BNN follows the configuration provided in \cite{FPBNN}. This network contains 6
convolutional layers, 3 pooling layers, and 3 fully-connected layers.
Pooling layers are computed in the same arrays as the preceding convolutional
layers. Data transfer between convolutional and pooling layers is considered
part of the computation phase since it is intra-array communication. The input
is a 32$\times$32 image with 3 channels. All filters are 3$\times$3 and all pooling layers are
2$\times$2 maxpool. There are 128 filters for the first two convolutional layers, 256
filters for the third and fourth, and 512 for the fifth and sixth. The output of
the last convolutional layer is 8192 neurons which are input to the first
fully-connected layer. There are two hidden layers of 1024 bits each and the
final output layer is 10 neurons. It achieves an accuracy of 86.31\% on the
CIFAR-10 dataset.

Convolutional FINN, the network in \cite{FINN}, is similar in structure. It also has 6
convolutional layers, 3 pooling layers, and 3 fully-connected layers. All filters are 3$\times$3 and all pooling layers are 2$\times$2
maxpool. There are 64 filters for the first two convolutional layers, 128
filters for the third and fourth, and 256 for the fifth and sixth. There are two
fully-connected hidden layers that each have 512 neurons. The output is 10 16-bit neurons. It
achieves and accuracy of 80.1\% on the CIFAR-10 dataset.

AlexNet, following Liang et al.\cite{FPBNN}, use the XNOR-NET \cite{xnornet} configuration to perform image classification on the ImageNet dataset. It achieves an accuracy of 42.90\% for top-1 and 66.80\% for top-5. The network configuration is similar to that of the other convolutional networks with the majority of the operations remaining binary. However, there are a few significant differences. The weights of some layers are not binary, in which case they are processed one bit at a time. Multiplications are introduced to perform scaling after each popcount operation. Batch-normalization is performed with multiplication, instead of the shift-based method presented earlier. Multiplications are performed inside the \arch\ array, requiring them to be performed bit-wise and thus have a relatively high latency. The network has 5 convolutional layers, 3 of which are followed by pooling layers. After convolution, there are three fully-connected layers and the final layer is 1000 neurons representing each of the 1000 classes. AlexNet is substantially larger than the previous networks, demanding a significant amount of memory. This is to be expected as it is too large to store on-chip for most hardware implementations; Liang et al.\cite{FPBNN} rely on external memory to store the network parameters.

We introduce a customized binarized neural network, BioNET, which finds if two
strings of genomic information match or not. 
It is
used as part of a larger genetic algorithm and it achieves an accuracy of
93.4\%, on average. 
The
core operations are the same as other BNNs, however, convolution can
occur in different dimensions for different layers. Additionally, the layer
sizes are non-uniform, which are listed in Table \ref{tab:knet}. As this network
is the first of its kind, there is no direct baseline for comparison. As
Liang et al.~\cite{FPBNN} is an FPGA specifically tailored for BNNs, which
significantly outperforms CPU and GPU alternatives, we use a BioNET
implementation on it as a baseline.
To this end, using reported layer sizes and characterization data
from \cite{FPBNN}, we extract a latency and energy model for 
%the performance of
an arbitrarily sized network on the FPGA.

\begin{table}[ht]
%\vspace{-.3cm}
\caption{BioNET (FC: fully connected; Conv: convolutional). \label{tab:knet}}
  \centering
  \resizebox{.7\linewidth}{!}{
%	
%\vspace{-.3cm}
\begin{tabular}{  c | c | c | c | c | c | c  }
%\hline
Layer & Type & Input & \# Filters & Filter Size & Pool Size & Output  \\
\hline
1 & Conv & 4x100x1 & 64 & 4x3 & 1x5 & 1x20x64 \\
%\hline
2 & Conv & 1x20x64 & 32 & 1x5x64 & 1x2 & 1x10x32 \\
%\hline 
3 & Conv & 1x10x32 & 20 & 1x4x32 & 1x2 & 1x5x20 \\
%\hline 
4 & FC & 100 & - & - & - & 40 (10-bit) \\
%\hline
\end{tabular}
}

\end{table}

\section{Evaluation}
\label{sec:eval}
%\b{
We evaluate \arch\ for both Modern (M) and Future (F) MTJs, per Table \ref{tab:mtj}. We restrict operations to only NAND, NOT, and COPY. We consider both an ideal (I) case, where there is no additional overhead to peripheral circuity, and one with estimates for peripheral circuitry (P). Additionally, we consider two different tiles sizes, 1024x1024 and 2048x2048. As data layout is generally more optimal in the 1T design and the 3T design incurs a large area overhead, we report results only for the 1T design. 
%}

\subsection{Single Inference Pass}
\noindent 
We start the evaluation with 
the performance and energy characterization for a single inference pass, which translates into the processing of a single image considering FP-BNN and FINN.
This reflects the time and energy required to write the input data, compute all
layers, and read final output values. 
Tables \ref{tab:totals_l} and \ref{tab:totals_e} show the total latency
and energy
for all networks considering different configurations,
along
with the latency and energy of baseline FPGA implementations. We show four configurations for future MTJ devices, with and without overhead for peripheral circuitry for tiles sizes of 1024 and 2048. We show a single configuration for modern MTJ devices, as they are less competitive. 

\arch\ typically has a lower latency than the FPGA for larger benchmarks but higher latency for smaller benchmarks. However, the main advantage of \arch\ is increased energy efficiency. It provides significant reductions in end-to-end energy consumption for all benchmarks when using future MTJ devices. Using modern MTJ devices, \arch\ has a higher, yet comparable, energy cost to the FPGAs. Our energy advantage decreases with larger benchmarks. This is because as benchmarks get larger there are more opportunities for parallelism. We opt to aggressively exploit these opportunities, providing the previously mentioned latency advantage, however this significantly increases our energy consumption.

Peripheral circuitry has a modest energy cost but adds a significant amount of latency. Much of the latency overhead comes from the addressing required to specify the rows and columns in which computation occurs, in addition to the addressing for memory operations. Fortunately, the peripheral latency for computation can be mitigated. The rows in which computation occur change infrequently, often separated by thousands of operations. Hence, the latency required to specify the rows can be amortized. Additionally, when a logic operation is performed, it can be applied to many rows, yet the peripheral latency for initiating the computation only has to be paid once.

Using smaller array sizes decreases latency but increases energy consumption. This is because smaller arrays have smaller rows and the computation must be divided into more rows. This increases parallelism and provides better performance, as computation in each row can proceed simultaneously. However, this also introduces additional communication overheads. Computation divided into multiple rows must be combined to produce a final result. Therefore, configurations with smaller tiles must perform more total operations for the same computation, which consumes more energy.

\begin{table}[th]
\caption{Overall latency (s). {NVIDIA Tesla K40 GPU takes 0.113s \cite{FPBNN}.}}
  \centering
  \resizebox{.8\linewidth}{!}{
%	
%\begin{tabular}{|c|c|c|}
\centering
\begin{tabular}{  c || c | c | c | c | c | c  }
%\hline
  & {\bf FPGA} & {\bf F-I-1024} & {\bf F-P-1024} & {\bf F-I-2048} & {\bf F-P-2048} & {\bf M-I-1024}  \\ 
 \hline \hline
 %Data goes here
FPBNN XNOR-Net &
%Latency FPGA
1.16e-3 \cite{FPBNN} & 
%F-I-1024
1.84e-4 & 
%F-P-1024
3.08e-4 & 
%F-I-2048
3.09e-4 & 
%F-P-2048
5.14e-4 & 
%M-P-1024
5.53e-4  \\
%XNOR-Net & \cite{FPBNN} & & & & & \\
\hline
FPBNN CIFAR-10 & 
%Latency FPGA
1.3e-4 \cite{FPBNN} & 
%F-I-1024
9.21e-5 & 
%F-P-1024
1.54e-4 & 
%F-I-2048
1.53e-4 & 
%F-P-2048
2.54e-4 & 
%M-P-1024
2.76e-4  \\
%CIFAR-10 & \cite{FPBNN}& & & & & \\
\hline
FPBNN FC & 
%Latency FPGA
3.39e-6 \cite{FPBNN} & 
%F-I-1024
5.05e-5 & 
%F-P-1024
8.48e-5 & 
%F-I-2048
9.34e-5 & 
%F-P-2048
1.54e-4 & 
%M-P-1024
1.52e-4  \\
%FC & \cite{FPBNN}& & & & & \\
\hline
FINN CIFAR-10 & 
%Latency FPGA
2.83e-4 \cite{FINN} & 
%F-I-1024
8.56e-5 & 
%F-P-1024
1.43e-4 & 
%F-I-2048
1.42e-4 & 
%F-P-2048
2.35e-4 & 
%M-P-1024
2.57e-4  \\
%CIFAR-10 & \cite{FINN} & & & & & \\
\hline
FINN FC & 
%Latency FPGA
2.44e-6 \cite{FINN} & 
%F-I-1024
3.80e-5 & 
%F-P-1024
6.29e-5 & 
%F-I-2048
7.33e-5 & 
%F-P-2048
1.21e-4 & 
%M-P-1024
1.14e-4  \\
%FC & \cite{FINN} & & & & & \\
%\hline
\end{tabular}
}
%\
\label{tab:totals_l}

\end{table}

\begin{table}[th]

\caption{ Overall energy (J).{NVIDIA Tesla K40 GPU takes 26.5J \cite{FPBNN}.}}

  \centering
  \resizebox{.8\linewidth}{!}{
%	
%\begin{tabular}{|c|c|c|}
\centering
\begin{tabular}{  c || c | c | c | c | c | c  }
%\hline
  & {\bf FPGA} & {\bf F-I-1024} & {\bf F-P-1024} & {\bf F-I-2048} & {\bf F-P-2048} & {\bf M-I-1024}  \\ 
 \hline \hline
%Data goes here
FPBNN XNOR-Net & 
%Energy FPGA
3.04e-2 \cite{FPBNN} & 
%F-I-1024
5.35e-3 & 
%F-P-1024
5.58e-3 & 
%F-I-2048
4.96e-3 & 
%F-P-2048
5.17e-3 & 
%M-P-1024
3.29e-1  \\
%XNOR-Net& \cite{FPBNN} & & & & & \\
\hline
FPBNN CIFAR-10 & 
%Energy FPGA
2.99e-4 \cite{FPBNN} & 
%F-I-1024
3.06e-5 & 
%F-P-1024
3.19e-5 & 
%F-I-2048
2.86e-5 & 
%F-P-2048
2.98e-5 & 
%M-P-1024
1.85e-3  \\
%CIFAR-10& \cite{FPBNN} & & & & & \\
\hline
FPBNN FC & 
%Energy FPGA
1.68e-5 \cite{FPBNN} & 
%F-I-1024
1.03e-6 & 
%F-P-1024
1.08e-6 & 
%F-I-2048
9.92e-7 & 
%F-P-2048
1.03e-6 & 
%M-P-1024
6.23e-5  \\
%FC& \cite{FPBNN} & & & & & \\
\hline
FINN CIFAR-10 & 
%Energy FPGA
5.34e-4 \cite{FINN} & 
%F-I-1024
9.49e-6 & 
%F-P-1024
9.90e-6 & 
%F-I-2048
9.17e-6 & 
%F-P-2048
9.57e-6 & 
%M-P-1024
5.75e-4  \\
%CIFAR-10& \cite{FINN} & & & & & \\
\hline
FINN FC & 
%Energy FPGA
1.45e-5 \cite{FINN} & 
%F-I-1024
1.46e-7 & 
%F-P-1024
1.52e-7 & 
%F-I-2048
1.76e-7 & 
%F-P-2048
1.83e-7 & 
%M-P-1024
8.86e-6  \\
%FC& \cite{FINN} & & & & & \\
%\hline
\end{tabular}
}

\label{tab:totals_e}

\end{table}

Using the same approach as for the previously described networks, we implement BioNET on \arch\ and report the latency and energy consumption. As BioNet is quite small, we expect the FPGA to have a significant latency advantage. Table \ref{tab:knetpl} lists the results.  Consistent with previous findings, the FPGA is faster but \arch\ is more energy efficient. As the target application, gene sequencing, is not real-time, throughput is a more significant metric than latency. In the next section we show how we can achieve high throughput while benefiting from \arch 's energy efficiency. 

\begin{table}[th]

\caption{Latency and energy characterization for BioNET.}
\label{tab:knetpl}

  \centering
  \resizebox{.45\linewidth}{!}{
    \centering
    \begin{tabular}{c|c|c|c}
         %& FP-BNN \cite{FPBNN} & \arch\ F-T & \arch\ F-T (FPC+N)\\
	  & {\bf FPGA} \cite{FPBNN} & {\bf F-I-1024} & {\bf F-P-1024}\\
        \hline
        Latency (s) & 9.95e-8 & 4.20e-5 & 7.04e-5 \\
        Energy (J) & 2.37e-7 & 1.09e-6 & 1.13e-6
    \end{tabular}
}

\end{table}

\subsection{Pipe-lining and Scaling}
\noindent 
In this section we show how the scalability of \arch\ can be used to drastically
increase performance, while still taking advantage of the inherent energy
efficiency of MTJ based computation. Since inference is performed across
multiple arrays, an opportunity for pipe-lining exists. 
Once data transfer
between different arrays is complete, computation in each array is independent.
This allows it to proceed in parallel. Given that \arch\ is energy efficient,
many additional arrays can be added to the neural network implementations while
maintaining a modest power budget. 

\begin{figure}[tph]

%
%\centering
   % \begin{minipage}[c][1\width]{0.49\textwidth}
    %\centering
%%%%%%%%%%%%%%%%%%%%%%%%%%%%%%%
\includegraphics[scale=0.23]{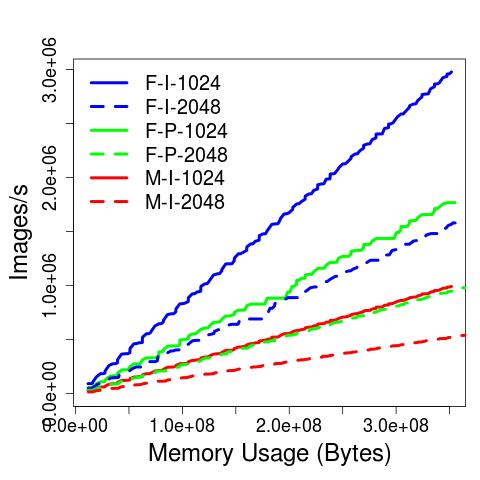}
\label{fig:FPBNN_FC_t}
%\hfill
\hspace{1cm}
\includegraphics[scale=0.23]{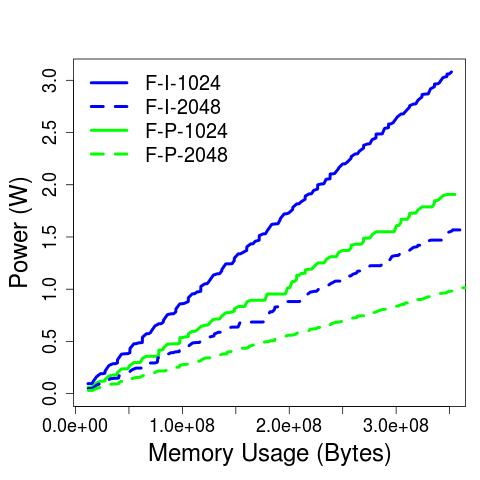}
\label{fig:FPBNN_FC_p}
\caption{
Fully-Connected MNIST FP-BNN classifier implemented in \arch. FP-BNN on FPGA \cite{FPBNN} has throughput of less than 1.56e6 images/s and consumes \SI{26.2}{\watt}. With modern MTJs, \arch\ consumes \SI{53.16}{\watt} when using 300 MB of memory.}
\label{fig:FPBNN_FC_b}
%%%%%%%%%%%%%%%%%%%%%%%%%%%%%%%

%\end{minipage}
%\hfill
%
%\begin{minipage}[c][1\width]{0.49\textwidth}
%%%%%%%%%%%%%%%%%%%%%%%%%%%%%%%
\includegraphics[scale=0.23]{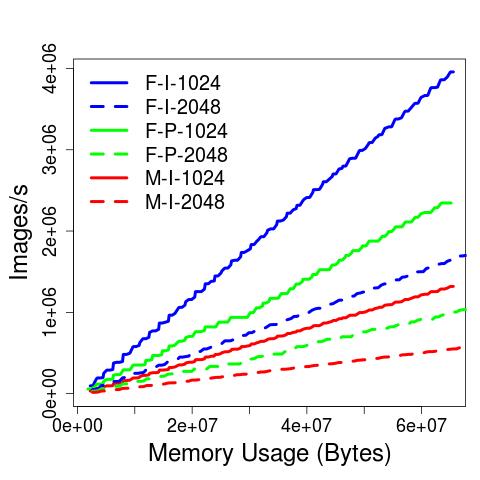}
\label{fig:FINN_FC_t}
%\hfill
\hspace{1cm}
\includegraphics[scale=0.23]{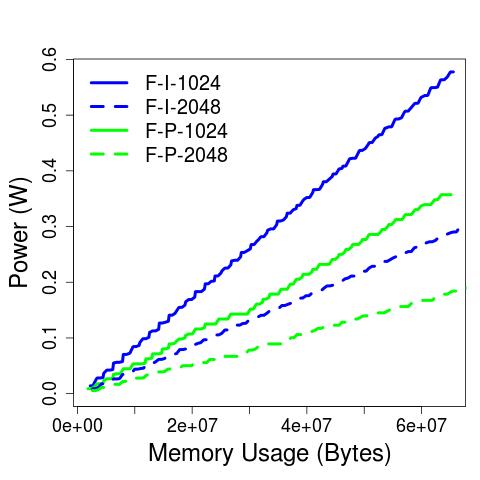}
\label{fig:FINN_FC_p}
\caption{
Fully-Connected MNIST FINN classifier implemented in \arch. FINN on FPGA \cite{FINN} has throughput of 1.56e6 images/s and consumes \SI{22.6}{\watt}. With modern MTJs, \arch\ consumes \SI{10.82}{\watt} when using 60 MB of memory.}
\label{fig:FINN_FC_b}

%%%%%%%%%%%%%%%%%%%%%%%%%%%%%%%
%\end{minipage}
%
%\vspace{-.3cm}
\end{figure}

\begin{figure}[tph]

%
%\centering
    %\begin{minipage}[c][1\width]{0.49\textwidth}
   % \centering
%%%%%%%%%%%%%%%%%%%%%%%%%%%%%%%
\includegraphics[scale=0.23]{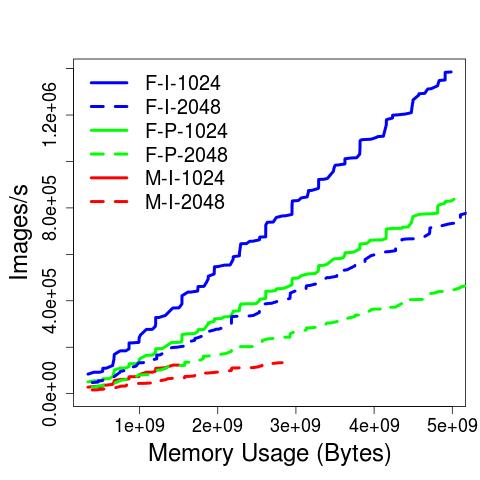}
\label{fig:FPBNN_cifar10_t}
%\hfill
\hspace{1cm}
\includegraphics[scale=0.23]{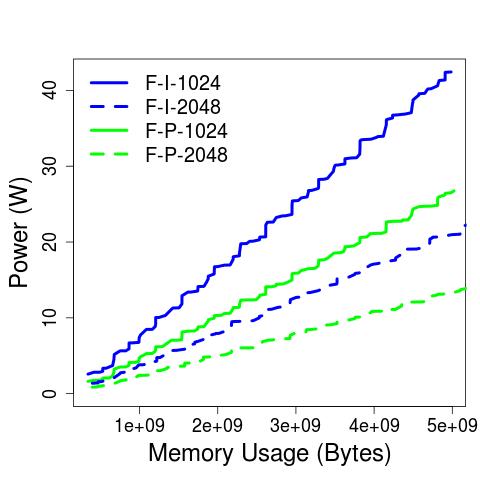}
\label{fig:FPBNN_cifar10_p}
\caption{
FP-BNN CIFAR-10 classifier implemented in \arch. FP-BNN on FPGA \cite{FPBNN} has throughput of 87.6e3 images/s and consumes \SI{26.2}{\watt}. With modern MTJs, \arch\ has a throughput of 123,225 images/s and uses 1.5 GB of memory when limited to \SI{235}{\watt}. 
}
\label{fig:FPBNN_cifar10_b}
%%%%%%%%%%%%%%%%%%%%%%%%%%%%%%%
%\end{minipage}
%\hfill
%
%\begin{minipage}[c][1\width]{0.49\textwidth}
%%%%%%%%%%%%%%%%%%%%%%%%%%%%%%%
\includegraphics[scale=0.23]{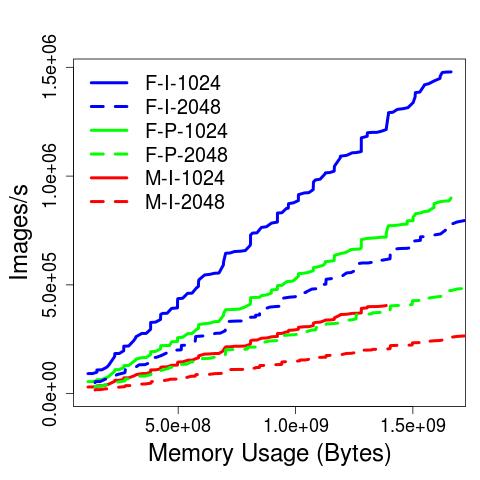}
\label{fig:FINN_cifar10_t}
%\hfill
\hspace{1cm}
\includegraphics[scale=0.23]{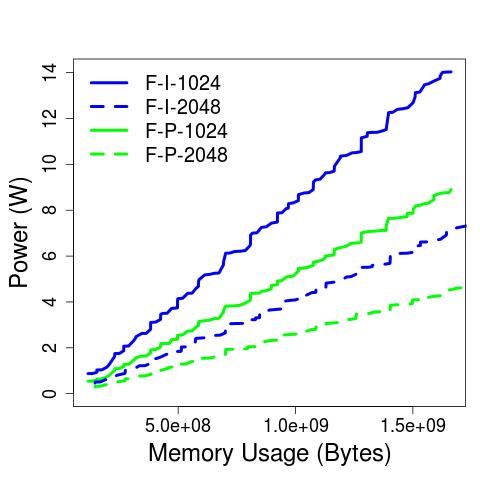}
\label{fig:FINN_cifar10_p}
\caption{
FINN CIFAR-10 classifier implemented in \arch. FINN on FPGA \cite{FINN} has throughput of 21.9e3 images/s and consumes \SI{11.7}{\watt}. With modern MTJs, \arch\ has throughput of 405,398 images/s and uses 1.39 GB of memory when limited to \SI{235}{\watt}.}
\label{fig:FINN_cifar10_b}

%%%%%%%%%%%%%%%%%%%%%%%%%%%%%%%
%\end{minipage}
%
%\vspace{-.3cm}
\end{figure}

\begin{figure}[tph]

%
%\centering
 %   \begin{minipage}[c][1\width]{0.49\textwidth}
 %   \centering
%%%%%%%%%%%%%%%%%%%%%%%%%%%%%%%
\includegraphics[scale=0.23]{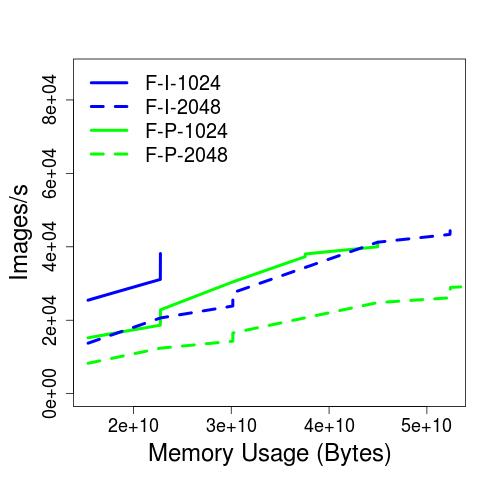}
\label{fig:AlexNet_t}
%\hfill
\hspace{1cm}
\includegraphics[scale=0.23]{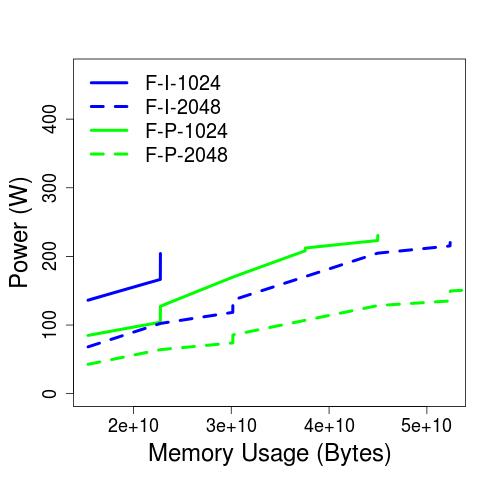}
\label{fig:AlexNet_p}
\caption{
AlexNet ImageNet classifier implemented in \arch. FPBNN FPGA \cite{FPBNN} has throughput of 863 images/s and consumes \SI{26.2}{\watt}. With modern MTJs, \arch\ has a throughput of 711.33 images/s and uses 796 MB of memory when limited to \SI{235}{\watt}.
}
\label{fig:AlexNet_b}
%%%%%%%%%%%%%%%%%%%%%%%%%%%%%%%

%\end{minipage}
%\hfill
%
%\begin{minipage}[c][1\width]{0.49\textwidth}
%%%%%%%%%%%%%%%%%%%%%%%%%%%%%%%
\includegraphics[scale=0.23]{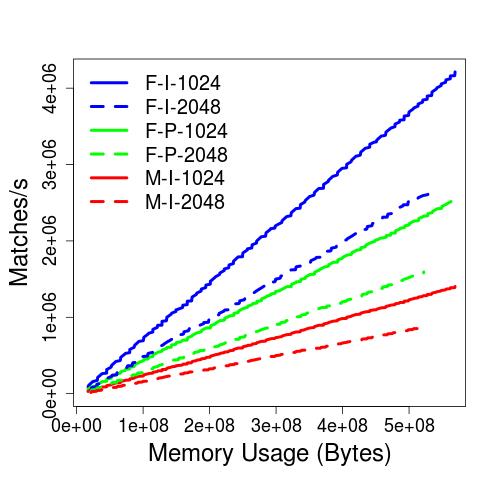}
\label{fig:knet_t}
%\hfill
\hspace{1cm}
\includegraphics[scale=0.23]{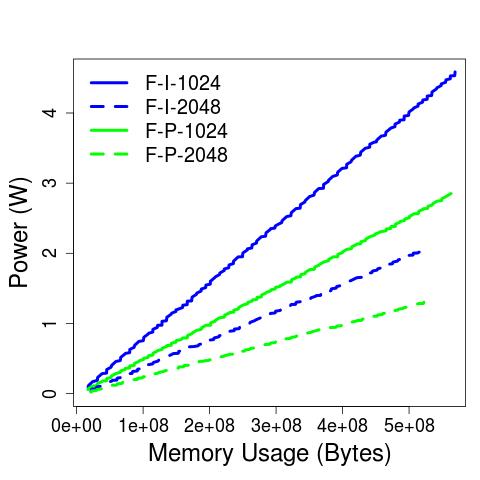}
\label{fig:knet_p}
\caption{BioNET implemented in \arch. FP-BNN FPGA \cite{FINN} consumes \SI{26.2}{\watt}. With modern MTJs, \arch\ consumes \SI{77.49}{\watt} when using 500 MB of memory.}
\label{fig:BioNET_b}

%%%%%%%%%%%%%%%%%%%%%%%%%%%%%%%
%\end{minipage}
%
%\vspace{-.3cm}
\end{figure}

\arch\ arrays are individual banks, many of which can reside on a single chip. Arrays
that are dedicated to consecutive layers can be placed near each other to
minimize the distance data must be transported. The cost of this communication
is the latency and energy required to read from the source array and to write to
the destination array. Since different banks can be accessed simultaneously, the
destination array can be written at the same time that the source array is
being read, nearly halving the latency. However, to be conservative in the
estimation of the communication overhead, we assume that these memory transfers
are entirely sequential. 

The base pipeline
configurations are those described in Sect.\ref{sec:setup}, with the same number of 
arrays dedicated for the convolutional networks {(CIFAR-10/ImageNet classifiers)} and for the fully-connected
networks {(MNIST classifiers)}. However, with pipe-lining, each array can be computing a layer for a
different input image at the same time. Effectively, convolutional networks are computed on a 9-stage pipeline, fully-connected networks on a 5-stage
pipeline, and AlexNet on an 8-stage pipeline. {Additional \arch\ arrays can then be repeatedly added to increase performance. } Each time an array
is added it is dedicated to the layer that will result in an increase in the
overall throughput the most. Adding \arch\ arrays is analogous to adding ALUs to
a traditional pipeline. 

{ Throughput and power are reported for increasingly larger configurations .} Throughput and power scale with the number of
arrays, however energy per image remains nearly constant. FP-BNN \cite{FPBNN}
does not report specific throughput numbers. Their CIFAR-10 classifier is
reported to have slightly less than 4$\times$ the throughput of the CIFAR-10 classifier
in FINN \cite{FINN}. Thus, we estimate the throughput of FP-BNN \cite{FPBNN} as 4$\times$ the throughput of FINN \cite{FINN}. {The MNIST classifier in FP-BNN \cite{FPBNN} has more neurons per layer than that in FINN \cite{FINN} and has additional non-binary operations. Thus, we assume that FINN provides an upper bound on the throughput of FP-BNN}. For XNOR-Net, they report a latency of \SI{1.16}{\milli \second}. A lower bound on their throughput would be 1/latency = 862 Images/sec. As XNOR-Net is too large to fit in the FPGA memory, we assume that this is a reasonable approximation.
%, and is used as comparison to the \arch\
%implementation. 
%For the MNIST classifier, FINN uses a smaller network
%configuration than FP-BNN. Thus, we compare the throughput of the \arch\ MNIST
%classifier with the FP-BNN topology to the throughput of FINN.

Fig.s \ref{fig:FPBNN_FC_b}-\ref{fig:BioNET_b} depict
throughput and power against memory usage for all benchmarks and configurations we considered for future MTJ devices. We report the ideal case (I), which only accounts for the latency and energy of the MTJs, and also report results accounting for the peripheral circuitry (P). This is done for both tiles sizes of 1024 and 2048. For throughput for modern MTJs, we only show the ideal case, as they are not as competitive. Due to their lack of extreme energy efficiency, power consumption for modern devices increases significantly with the introduction of pipe-lining, and thus is omitted from the figures. The \arch\ XNOR-Net configuration is highly optimized for performance, and thus consumes significant amounts of power, even for future MTJs. We increase performance from the base configuration until the power consumption reaches \SI{235}{\watt}, the same power consumption as the GPU implementation \cite{FPBNN}. Such aggressive performance optimization is not possible with modern MTJs, therefore we opt to reduce performance to keep the modern power budget within reason. Overall, we observe that \arch\ can provide very high throughput while remaining at a very modest power budget. For all networks other than XNOR-Net, \arch's power budget is well below that of the FPGAs. XNORNet represents the extreme case where we heavily optimized for latency and throughput at the cost of energy efficiency. As a result, even the smallest configuration consumes a considerable amount of power. We believe that optimizing for latency is more appropriate for XNOR-Net, as the memory requirements are currently beyond the scope of edge, low-power devices. Thus, we choose to provide very high performance and consume moderate amounts of power. 

\begin{wrapfigure}{r}{0.45\textwidth}

%
%\begin{minipage}[c][1\width]{0.5\textwidth}

\centering
%%%%%%%%%%%%%%%%%%%%%%%%%%%%%%%
\includegraphics[width=0.22\textwidth]{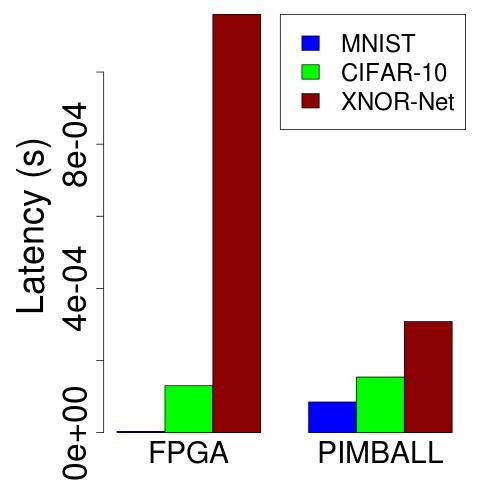}
\label{fig:BenchMarkLatencies}
\hfill
\includegraphics[width=0.22\textwidth]{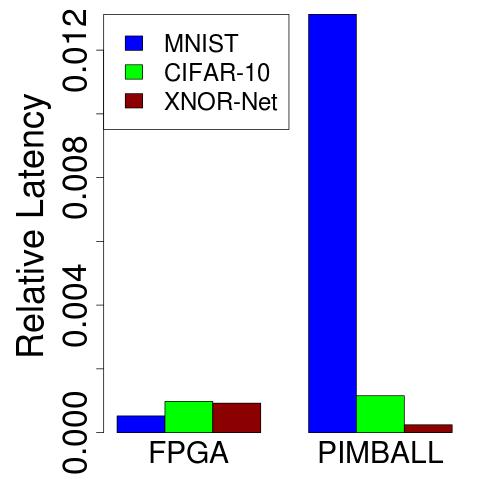}
\label{fig:RelativeBenchMarkLatencies}

\label{fig:AlexNet_b}
%%%%%%%%%%%%%%%%%%%%%%%%%%%%%%%

%\end{minipage}
%

\caption{
Latencies of each benchmark on Stratix V FPGA \cite{FPBNN} and \arch\ (F-P-1024) (a). Relative latency of  \arch\ to NVIDIA Tesla K40 GPU (b).
}
\label{fig:Latencies}

\end{wrapfigure}

While of widely varying sizes, each of the networks demonstrate similar trends. As expected, smaller tile sizes demonstrate higher throughput and higher power consumption. For a given memory usage, the configurations which consider peripheral circuitry actually have a lower power consumption. This is because peripheral circuitry has a larger effect on latency than energy. 
%The increases latency due to peripheral circuitry slows down the computation, and as a result, reduces the performance and power.
 Notice also that as BioNET is not a stand-alone application but a kernel inside of a 
larger genomics algorithm, throughput is more valuable than latency. 
%As we can see in Fig.\ref{fig:BioNET_b}, \arch\ can achieve a high throughput at a very low power budget. 

%The impact of peripheral circuitry, however, is mitigated by the access patterns involved in computation within the array. During communication and the initiation of computation, the cost of peripheral circuitry is high, due to the specification of row addresses. However, the number of active rows does not need to be changed for long periods of time during computational phases. Additionally, the latency and energy cost associated with applying voltages to the bitlines is quite high relative to that of MTJ switching. However, each bitline is shared by many MTJs (and logic operations) and thus this cost is amortized. The peripheral circuitry has less impact during these phases. 

Putting it all together, the most noticeable advantage of \arch\ is its energy
  efficiency. This makes it an ideal candidate for mobile applications where
  power consumption is critical. For example, the FP-BNN CIFAR-10 classifier
  which achieves an accuracy of 86.31\% -- accounting for predicted future
peripherals and reliable NAND only operations -- could process 60 Images/s at a power
budget of \SI{1.9}{\milli \watt}. However, \arch\ is not restricted to such
target applications. If high performance is desired, it can easily be scaled up
to achieve state-of-the-art throughputs. Even at such throughputs, the power
consumption remains only a few watts. 
Such scalability is another key advantage
of \arch. By doing computation in memory, large penalties for data transfers
are mitigated. The capacity allows for a large number of parameters
to be stored entirely on chip. These combine to make \arch\ particularly well
suited for large scale applications which operate on large volumes of data. This
is reflected in the fact that \arch\ does not provide much benefit over an FPGA for a single image
inference, but is capable of a much higher throughput. This scalability can also be seen in Fig.\ref{fig:Latencies}. The latency of \arch\ does not increase dramatically with benchmark size (as we move from MNIST to XNOR-Net), whereas the FPGA latency \cite{FPBNN} does. Additionally, the relative latency of \arch\ to an NVIDIA Tesla K40 GPU decreases with increasing benchmark size, suggesting better scalability.

\section{Related Work}
\label{sec:rel}

Numerous NN accelerators have been proposed for forward propagation on traditional compute substrates. For example, DaDianNao
\cite{dadiannao} uses a multi-chip system to implement high-precision networks.
Eyeriss \cite{eyeriss} develops a data flow to improve the energy efficiency on a 
%(fabricated)
spatial architecture. 
%\b{
Tetris \cite{gao2017tetris} and Neurocube \cite{kim2016neurocube} utilize 3D stacked memories. Neurocube uses the Hybrid Memory Cube (HMC) from Micron and performs the computation within its logic layer. Programmable state machines within the HMC are responsible for data movement and program flow. Tetris \cite{gao2017tetris} also uses the HMC but focuses on using the 3D memory to  optimize other components. By utilizing the high throughput of the memory and moving some of the computations to the near-memory logic layers, the authors were able to dedicate more area to computational units and alleviate overheads due to data transfer. While computation still occurs external to the memory, Tetris and Neurocube reveal the bottleneck that data transfer imposes and how creating a memory-centric design can mitigate it. 
%}
%and demonstrated a fabricated chip. 
Many FPGA accelerators also exist \cite{high,godeeper, throughput}.
FPGA accelerators capitalizing on the benefits of binarization form
the most relevant subset of this rich body of work, including 
\cite{SOFTPROGRAM, FINN, FPBNN} which can achieve
%These networks achieved 
significantly higher throughput and energy efficiency
than competing CPU and GPU based implementations. {Methods to exploit redundancy in BNNs to increase performance on FPGAs was presented in \cite{fu2018towards}}. SRAM based acceleration is proposed in \cite{neuralcache}. However, by definition, SRAM cannot support large data sets. {A software-reconfiguralbe accelerator for binarized neural networks intended for use with micro-controller systems was proposed in \cite{conti2018xnor}. They can implement small networks on-chip but must rely on external memory for practical networks. A recent paper \cite{kim2019nand} modifies XNOR-Net to use NAND operations instead and implements accelerators in DRAM and SRAM. A hybrid design was proposed in \cite{lee2017energy}, where they trade off accuracy for increased energy efficiency by implementing a significant portion of the network with stochastic computing. }
%\b{
The authors of \cite{nurvitadhi2016accelerating} implemented fully-connected layers of BNNs on multiple platforms, including a CPU, GPU, FPGA, and ASIC. They were able to demonstrate quantitative  benefits of batching and binarization and improvements due to hardware acceleration. 
%}

When it comes to  in- or near-memory NN acceleration using non-volatile memory, RRAM represents the most common substrate:
%Neural nets have also been
%implemented in RRAM. 
%As representative examples, a convolutional MNIST classifier was proposed in
%\cite{rram2016}; a LeNET model on MNIST and AlexNet on ImageNet for a %variety of bit
%precisions, in \cite{rram2017}; binarized
%fully-connected layers, in \cite{rram2018}.
%also uses an RRAM accelerator to implement  
%\cite{rram2018} proposes an architecture to implement 
%Further comparisons to RRAM are presented in Section
%\ref{sec:rram}. 
%\subsection{Comparison to RRAM}
%\label{sec:rram}
%As RRAM accelerators have been used for similar applications, we make some additional comparisons to \arch. 
RRAM, like STT-MRAM, is a resistive memory technology, where the state of a cell is stored in the resistivity of the material. Notably, RRAM has multiple resistive states and the extreme ends demonstrate a higher TMR than STT-MRAM. However, RRAM suffers from degradation with use. The state can be switched only a limited number of times before the device begins to fail. 

A few reduced precision and binary neural networks have been implemented in RRAM. These networks usually take on a different form than the structure proposed in this work. RRAM accelerators typically store the weights of a network in a crossbar. The inputs (neuron values) are the voltages applied to the wordlines. The outputs of the operations are the currents on the bitlines, which are sent to an ADC (analog to digital converter) for multi-bit precision networks or a sense amplifier for binary networks. Most implementations rely on external digital logic circuits for a significant amount of the computation, such as the addition and thresholding. Thus, the RRAM crossbar is typically used just as an accelerator for the matrix-vector multiplications. This is 
%contrary to 
in stark contrast with 
\arch\, where all operations are performed within the memory array itself. 

As representative examples, convolutional MNIST classifiers were proposed in
\cite{rram2016} and \cite{rram2016b}.
%; a LeNET model on MNIST and AlexNet on ImageNet for a variety of bit
%precisions, in \cite{rram2017}; binarized
%fully-connected layers, in \cite{rram2018}.
%An RRAM accelerator was proposed \cite{rram2016b} where convolutional networks were used to classify the MNIST dataset. 
In \cite{rram2016b}, the authors binarized the output from the RRAM crossbar to avoid using ADCs and used inputs as selection signals to avoid using DACs (digital to analog converters). This was significant as ADCs and DACs typically consume the majority of energy and power in convolutional RRAM networks. Many networks were implemented this way, where the most energy efficient network (that was as accurate as the networks evaluated in our study) achieved an accuracy of 98.47\% at \SI{2.58}{\micro \joule/Image}. 
A similar approach was also used in \cite{rram2017} to classify MNIST and ImageNet data. 

On the other hand, the design from \cite{rram2018} implements the fully-connected layers of a convolutional neural network. They activate multiple wordlines simultaneously, use two columns storing complementary weights, and a sense amplifier to compare the difference in current between two bitlines. This allows multiplication, pop-count, and thresholding to occur at the same time within the array, and removes the need for an external adder and comparator. Their network has two 64 neuron layers followed by a 10 neuron output layer. 
%While covering only fully-connected layers, 
A \arch\ implementation of this same network would be slightly more energy efficient, if peripheral circuitry overheads could be minimized.
%is not taken into account. 
Considering practical overheads due to modern peripherals, 
%Considering the overhead incurred by practical peripheral circuitry, 
however, \arch\ becomes %significantly 
less energy efficient
%In both cases \arch\ would be 
and considerably slower. 
%Thus, the latency and energy efficiency of this analog based %computing generally outperforms \arch\. 

That said, covering only fully-connected layers, the advantage of this RRAM  design \cite{rram2018} can only exist for small networks. Increasing the input size degrades the accuracy.
%increases the error. 
Scaling to the networks evaluated in our study, which are multiple orders of magnitude larger, would make a direct implementation of 
%directly implementing 
the analog computing approach of this RRAM design impossible. 
%While this RRAM solution shows great potential for small NN, 
At the same time, it is subject to errors and noise due to the analog nature of computing, which may not always be masked by the implicit noise tolerance of NN. In fact, as recent work \cite{feinberg2018making} has demonstrated, NN on similar RRAM arrays may suffer from significant accuracy loss due to  noise.
%At the same time, as recent work has demonstrated\cite{feinberg2018making},  noted that RRAM accelerators often introduce significant levels of error due to their analog nature. 
%They developed error correcting codes that can be used with RRAM %accelerators to regain accuracy.

%We implement the same network size on \arch\ and compare the results in Table \ref{tab:rram2018}. As can be seen, \arch\ has a considerably higher latency than RRAM. This is mostly due to the fact that the RRAM implementation performs the computation with an analog comparison performed in one operation, whereas \arch\ performs a sequence of operations using digital summation, which takes longer. While these results would seem to indicate RRAM is superior, it is important to consider scalability. The RRAM implementation will not scale well to larger input sizes as differences between input vectors will be harder to distinguish as the input size increases. The digital approach of \arch\ enables it to handle any input size, the networks simulated in this paper have layers with thousands of neurons. The energy efficiencies of \arch\ and RRAM are comparable.

%Don't know if we want to include this table
%could be misleading as this is not fair comparison
%\begin{table}
%\begin{tabular}{ | c | c | c | c | }
%\hline
%Performance & P-BNN (RRAM) & \arch\ F (FP) & \arch\ F \\
%\hline
%Latency (ns) & 10.42 & 3003 & 1534 \\
%\hline
%Energy (pJ) & 34.48 & 47.12 & 29.08 \\
%\hline
%\end{tabular}
%\caption{Comparison between RRAM and \arch\ fully-connected layers}
%\label{tab:rram2018}
%\end{table}

Other spintronic substrates featuring computation capability within the memory array also exist: 
Pinatubo \cite{pinatubo} proposes an architecture to do general
bit-wise operations in non-volatile memory. The authors in \cite{angizi2018imce} use an MTJ subarray as part of an accelerator for low bit-width convolutional NN.  Recent work also covers multi-level MRAM cells to implement BNN \cite{pan2018multilevel}. Networks with binary weights utilizing STT-MRAM were proposed in \cite{angizi2019parapim}. Contrary to \arch, all of these platforms use sense amplifiers to perform computation. 
%By construction, the outputs of logic operations are immediately stored %within the array.  
In addition, all use additional digital logic circuity. Pinatubo \cite{pinatubo} embeds digital logic circuits into the memory for inter-subarray computation. The authors in \cite{pan2018multilevel} use an auxiliary processing unit to perform batch-normalization, multiplication, and pooling. The design in \cite{angizi2018imce} performs operations such as bit counting, summation, 
%activation function, 
quantization and batch normalization external to the array. This is contrary to \arch, which contains no additional digital logic circuitry and which performs all layers of the network entirely within the memory array. 
%\blueHL{
Another recent paper \cite{gupta2018felix}  proposes 
%single-cycle 
MTJ based computation without sense amplifiers in a crossbar topology. As opposed to \cite{gupta2018felix}, \arch\ can support true PIM semantics at scale: effectively, a \arch\ array is not any different than STT-MRAM when not used for computation. %Additionally, they use extra transistor between some of the rows to isolate regions of the array and increase parallelism. 
%A crossbar architecture is not suitable for \arch\ as it is intended to be capable of fitting into the memory heirarchy and takes a form much more %similar to traditional STT-MRAM arrays.}
%\textcolor{red}{
Pinatubo \cite{pinatubo} is not explicitly tailored to NN acceleration. At the same time the convolutional networks implemented in our study are an order of magnitude larger than the network proposed in \cite{pan2018multilevel}. 
Our baseline spintronic PIM substrate, CRAM, was introduced in \cite{cram}, and evaluated for simple and very small scale (non-binary) NN (limited to a single-neuron digit recognizer and 2D convolution, specifically) in \cite{masoud}. As we cover in Sect.\ref{sec:cram}, this basic memory cell and array structure cannot support BNN acceleration without modification. {A review of neuro-inspired computing on non-volatile memories is provided in \cite{yu2018neuro}.}

\section{Conclusion}
\label{sec:conc}
%\textcolor{red}
\noindent
{In this paper, we explored the design space of binary neural network (BNN) (forward propagation) acceleration on a spintronic processing in memory (PIM) substrate. The result is a scalable, high throughput, and specifically, highly energy efficient solution, \arch. 
%evaluated the performance of computational RAM
%(CRAM), an STT-MRAM based memory and computational substrate, on binary neural
%network forward propagation. %
We demonstrated that \arch\ can efficiently perform all core building blocks required for BNN forward propagation such as XNOR, pop-count, batch normalization, and thresholding, {\em entirely} within \arch\ arrays, without any need for external circuitry, be it analog  or digital, to offload computation.
%No external digital circuits
%are required. 

Our analysis revealed that the key strength of \arch\ is energy efficiency.
%Simulation results demonstrated that \arch\ is extremely energy
%efficient. It was found that 
A single inference pass on \arch\ can take significantly longer when compared to representative, FPGA-based hardware alternatives which belong to best performing solutions in this context. 
%higher than the corresponding FPGA implementations. 
However, achieving competitive and even better levels of throughput performance in \arch, while preserving the energy efficiency gains, is straight-forward. This is because \arch\ lends itself very well to array-level parallelism -- and pipe-lining computation with (relatively slower) communication -- in addition to the inherent intra-array row-level parallelism.
%\arch\ features inherent row-level parallelism of the \arch\ substrate,
%however, this can be
%compensated for via parallelism to achieve a high throughput. If a sufficient
We demonstrated that a practically feasible number of \arch\ arrays can outperform the throughput of the FPGA counterparts while consuming only a fraction of their power. 
%number of CRAM arrays are used, \arch\ can achieve throughputs higher than start
%of the art customized FPGAs and operate at a fraction of the power. 
This makes
\arch\ a candidate for both low power mobile and high performance computing applications.} 

Another key contribution of this study is novel memory cell and array architecture designs for spintronic PIM, which we incorporated in \arch\ arrays, and which enabled efficient BNN processing. \arch\ arrays are implicitly re-configurable, as any row in the array can participate in any kind of computation, according to the algorithmic needs of the underlying problem. Therefore, the presented \arch\ designs can also be expanded to other acceleration problems similar in nature (in terms of computation and communication characteristics) to BNN.

\bibliographystyle{ACM-Reference-Format}
\bibliography{ref}

\end{document}